\documentstyle[aps,preprint,eqsecnum,floats,epsf]{revtex}
\tighten
\draft
\begin{document}

\preprint{}
\title{Supersymmetric and Shape-Invariant Generalization for 
Nonresonant and Intensity-Dependent Jaynes-Cummings Systems}
\author{A.~N.~F. Aleixo\thanks{Electronic address:
        {\tt aleixo@nucth.physics.wisc.edu}}}
\address{Department of Physics, University of Wisconsin\\
         Madison, Wisconsin 53706 USA\thanks{{\tt Permanent address}:
Instituto de F\'{\i}sica, Universidade Federal 
         do Rio de Janeiro, RJ - 
         Brazil.}}
\author{A.~B. Balantekin\thanks{Electronic address:
        {\tt baha@nucth.physics.wisc.edu}},}
\address{Max-Planck-Institut f\"ur Kernphysik,
Postfach 103980, D-69029 Heidelberg, Germany
\thanks{{\tt Permanent address:}Department of Physics, University of Wisconsin,
         Madison, Wisconsin 53706 USA}}
\author{M.~A. C\^andido Ribeiro\thanks{Electronic address:
         {\tt macr@df.ibilce.unesp.br}},}
\address{Departamento de F\'{\i}sica - Instituto de Bioci\^encias,
         Letras e Ci\^encias Exatas\\
         UNESP, S\~ao Jos\'e do Rio Preto, SP - Brazil}

\date{\today}
\maketitle

\begin{abstract}
  A class of shape-invariant bound-state problems which represent
  transition in a two-level system introduced earlier are generalized
  to include arbitrary energy splittings between the two levels as
  well as intensity-dependent interactions. We show that the
  couple-channel Hamiltonians obtained correspond to the
  generalizations of the nonresonant and intensity-dependent
  nonresonant Jaynes-Cummings Hamiltonians, widely used in quantized
  theories of laser. In this general context, we determine the
  eigenstates, eigenvalues, the time evolution matrix and the
  population inversion matrix factor.
\end{abstract}


\section{Introduction}

The integrability condition called shape-invariance originates in
supersymmetric quantum mechanics \cite{Witten:1981nf,Cooper:1995eh}. 
The separable
positive-definite Hamiltonian $\hat H_1 = \hat A^\dagger  \hat A$ is
called shape-invariant if the condition
\begin{equation}
\hat A(a_1) \hat A^\dagger(a_1) =\hat A^\dagger (a_2)  \hat A(a_2) +
R(a_1) \,,
\label{eqsi}
\end{equation}
is satisfied \cite{Gendenshtein:1983vs}.  
In this equation $a_1$ and $a_2$ represent
parameters of the Hamiltonian. The parameter $a_2$ is a function of
$a_1$ and the remainder $R(a_1)$ is independent of the dynamical
variables such as position and momentum.  Even though not all
exactly-solvable problems are shape-invariant \cite{Cooper:1987tz}, 
shape invariance, especially in its algebraic formulation
\cite{Balantekin:1998mg,Chaturvedi:1998hi,Balantekin:1999wj}, 
has proven to be a powerful technique to study exactly-solvable
systems.  

In a previous paper \cite{Aleixo:2000bd} we used shape-invariance 
to calculate the energy eigenvalues and eigenfunctions for the
Hamiltonian  
\begin{equation}
\label{oldham}
\hat {\bf H} = \hat A^\dagger\hat A + {1\over 2}\left[\hat A,\hat
A^\dagger\right]\left(\hat\sigma_3+1\right) +
\sqrt{\hbar\Omega}\left(\hat\sigma_+\hat A+\hat\sigma_-
\hat A^\dagger\right)\,,
\label{eqjca}
\end{equation} 
where 

\begin{equation}
\hat\sigma_\pm = {1\over 2}\left( 
\hat\sigma_1\pm i\hat\sigma_2\right)\,,
\label{eqsig}
\end{equation}
and $\hat\sigma_i$, with $i=1,\,2,\, {\rm and}\,\, 3$, are the Pauli
matrices.

This is a generalization of the Jaynes-Cummings Hamiltonian
\cite{ref18}. A different, but related problem was considered in
Ref. \cite{ref18a}.  Our goal in this paper is to study further
generalizations of the Jaynes-Cummings Hamiltonian, first by
introducing a term proportional to $\sigma_3$ with an arbitrary
coefficient (the so-called nonresonant limit) and then by taking into
account the dependence of the coupling on the intensity of the field
(the so-called intensity-dependent nonresonant limit). In addition
to the energy levels we study the time evolution and the population
inversion matrix factor.

Introducing the similarity transformation that replaces $a_1$ with
$a_2$ in a given operator
\begin{equation}
\hat T(a_1)\, \hat O(a_1)\, \hat T^\dagger(a_1) = \hat O(a_2)
\label{eqsio}
\end{equation}
and the operators
\begin{equation}
\hat B_+ =  \hat A^\dagger(a_1)\hat T(a_1)
\label{eqba}
\end{equation}
\begin{equation}
\hat B_- =\hat B_+^\dagger =  \hat T^\dagger(a_1)\hat A(a_1)\,,
\label{eqbe}
\end{equation}
the condition of Eq. (\ref{eqsi}) can be written as a commutator
\cite{Balantekin:1998mg}
\begin{equation}
[\hat B_-,\hat B_+] =  \hat T^\dagger(a_1)R(a_1)\hat T(a_1)  \equiv
R(a_0)\,,
\label{eqcb1}
\end{equation}
where we used the identity
\begin{equation}
R(a_n) = {\hat T}(a_1)\,R(a_{n-1})\,{\hat T}^\dagger (a_1)\,,
\label{eqran}
\end{equation}
valid for any $n$.  The ground state of the Hamiltonian $\hat H_1 =
\hat A^\dagger \hat A = \hat B_+ \hat B_-$ satisfies the condition
\begin{equation}
\hat A\,\mid \psi_0\rangle = 0 = \hat B_-\,\mid\psi_0\rangle\,;
\label{eqaps0}
\end{equation}
and the unnormalized $n$-th excited state is given by 
\begin{equation}
\mid \psi_n\rangle \sim \left( \hat B_+ \right)^n \mid\psi_0\rangle
\label{eqpsn}
\end{equation}
with the eigenvalue
\begin{equation}
{\cal E}_n =  \sum_{k=1}^n R(a_k)\,. 
\label{eqen}
\end{equation}
We note that the Hamiltonian of Eq. (\ref{oldham}) can also be
written as 
\begin{equation}
\hat {\bf H} = \left[ \matrix{\hat T & 0\cr 0 & \pm 1 \cr}\right]\hat {\bf
h}_{\pm} \left[ \matrix{\hat T^\dagger & 0\cr 0 & \pm 1 \cr}\right]\,,
\end{equation}
where
\begin{equation}
\hat {\bf h}_{\pm} =  \hat B_+ \hat B_- + {1\over 2} R(a_0)
\left(\hat\sigma_3+1\right) \pm \sqrt{\hbar\Omega}\left(\hat\sigma_+
\hat B_- +\hat\sigma_- \hat B_+  \right) \,. 
\end{equation} 

\section{The Generalized Nonresonant Jaynes-Cummings Hamiltonian}

The standard Jaynes-Cummings model, normally used in quantum optics,
idealizes the interaction of matter with electromagnetic radiation by a
simple Hamiltonian of a two-level atom coupled to a single bosonic
mode\cite{ref21,ref22,Chaichian:1990ic,ref24,ref25,ref26}. This
Hamiltonian has a 
fundamental importance to the field of quantum optics and it is a
central ingredient in the quantized description of any optical system
involving the interaction between light and atoms. The Jaynes-Cummings
Hamiltonian defines a {\it molecule}, a composite system formed from
the coupling of a two-state system and a quantized harmonic
oscillator. In this case, its nonresonant expression can be  written
as 
\begin{equation}
\hat {\bf H} = \hat A^\dagger\hat A + {1\over 2}\left[\hat A,\hat
A^\dagger\right]\left(\hat\sigma_3+1\right) +
\alpha \left(\hat\sigma_+\hat A+\hat\sigma_-
\hat A^\dagger\right) +\hbar\Delta\,\hat\sigma_3\,,
\label{eqjcnr}
\end{equation}
where $\alpha$ is a constant related with the coupling strength and
$\Delta$ is a constant related with the detuning of the system. 

However, the harmonic oscillator systems, used in this context, is
only the simplest example of supersymmetric and shape-invariant
potential. Our goal here is to generalize that Hamiltonian for all
supersymmetric and shape-invariant systems. With this purpose and
following Ref. \cite{Aleixo:2000bd} we introduce the operator
\begin{equation}
\hat {\bf S} = \hat\sigma_+\hat A + \hat\sigma_-\hat A^\dagger\,,
\label{eqso}
\end{equation}
where the operators $\hat A$ and $\hat A^\dagger$ satisfy the shape
invariance condition of Eq.~(\ref{eqsi}). Using this definition we can
decompose the nonresonant Jaynes-Cummings Hamiltonian in the form 
\begin{equation}
\hat {\bf H} = \hat {\bf H}_o + \hat {\bf H}_{int}\,,
\label{eqjcnr1}
\end{equation}
where 
\begin{mathletters}
\label{eqhint}
\begin{eqnarray}
& & \hat {\bf H}_o = \hat {\bf S}^2\,,\\ & & \hat {\bf H}_{int} =
\alpha \,\hat {\bf S} + \hbar\Delta\,\hat\sigma_3\,. 
\end{eqnarray}
\end{mathletters}
First, we search for the eigenstates of $\hat {\bf S}^2$. In this case
it is more convenient to work with its $B$-operator expression, which
can be written as \cite{Aleixo:2000bd} 
\begin{equation}
\hat {\bf S}^2 = \left[ \matrix{\hat T & 0\cr 0 & \pm 1 \cr}\right]
\left[ \matrix{\hat B_-\hat B_+ & 0\cr 0 & \hat B_+\hat B_-
\cr}\right] \left[ \matrix{\hat T^\dagger & 0\cr 0 & \pm 1
\cr}\right] \equiv \left[ \matrix{\hat H_2 & 0\cr 0 & \hat H_1 
\cr}\right]\,, 
\label{eqso2}
\end{equation}
where $\hat H_2 = \hat T \hat B_- \hat B_+  \hat T^\dagger$.
Note the freedom of sign choice in Eq.~(\ref{eqso2}), which results in two
possible decompositions of $\hat {\bf S}^2$.  Next, we introduce the
states 
\begin{equation}
\mid \Psi^{(\pm)}\rangle =  \left[ \matrix{\hat T & 0\cr 0 & \pm 1
\cr}\right]  \left[ \matrix{C_m^{(\pm)}\mid m\rangle\cr
C_n^{(\pm)}\mid n\rangle \cr}\right]\,,
\label{eqps+-}
\end{equation}
where \ $C_{m,n}^{(\pm)} \equiv C_{m,n}^{(\pm)} \left[R(a_1), R(a_2),
R(a_3), \dots\right]$ \ are auxiliary coefficients and, $\mid
m\rangle$ and  $\mid n\rangle$ are the abbreviated notation for the
states $\mid \psi_m\rangle$ and $\mid \psi_n\rangle$ of
Eq.~(\ref{eqpsn}). Using Eqs.~(\ref{eqcb1}), (\ref{eqso2}) and
(\ref{eqps+-}), the commutation between $\hat H_1$ and a function of
$R(a_k)$, and the $\hat T$-operator unitary condition, we get 
\begin{eqnarray}
\hat {\bf S}^2\mid \Psi^{(\pm)}\rangle &=&  \left[ \matrix{\hat T &
0\cr 0 & \pm 1 \cr}\right] \left[ \matrix{\hat B_+\hat B_- + R(a_0) &
0\cr 0 & \hat B_+\hat B_- \cr}\right] \left[ \matrix{C_m^{(\pm)}\mid
m\rangle\cr C_n^{(\pm)} \mid n\rangle \cr}\right]  \nonumber\\ &=&
\left[ \matrix{\hat T & 0\cr 0 & \pm 1 \cr}\right] \left[
\matrix{{\cal E}_m + R(a_0) & 0\cr 0 & {\cal E}_n \cr}\right]\left[
\matrix{C_m^{(\pm)}\mid m\rangle\cr  C_n^{(\pm)}\mid n\rangle
\cr}\right]\,. 
\label{eqmsop2}
\end{eqnarray}
And using Eqs.~(\ref{eqran}) and (\ref{eqen}) we can write
\begin{eqnarray}
\hat T\left[ {\cal E}_m + R(a_0)\right]\hat T^\dagger &=&  \hat
T\left[ R(a_1) + R(a_2) + \cdots + R(a_m) + R(a_0)\right]\hat
T^\dagger \nonumber\\ &=&  R(a_2) + R(a_3) + \cdots + R(a_{m+1}) +
R(a_1) = {\cal E}_{m+1}\,. 
\label{eqtemt}
\end{eqnarray}
Hence the states 
\begin{equation}
\mid \Psi_m^{(\pm)}\rangle =  \left[ \matrix{\hat T & 0\cr 0 & \pm 1
\cr}\right]  \left[ \matrix{C_m^{(\pm)}\mid m\rangle\cr
C_{m+1}^{(\pm)}\mid m+1\rangle \cr}\right], \qquad m=0,1,2, \cdots 
\label{eqpsm+-}
\end{equation}
are the normalized eigenstates of the operator $\hat {\bf S}^2$
\begin{equation}
\hat {\bf S}^2\mid \Psi_m^{(\pm)}\rangle = {\cal E}_{m+1}\mid
\Psi_m^{(\pm)}\rangle\,. 
\label{eqs2psm}
\end{equation}
We observe that the orthonormality of the wavefunctions  imply in the
following relations among the $C$'s 
\begin{mathletters}
\label{eqcpm}
\begin{eqnarray}
\langle\Psi_m^{(\pm)}\mid \Psi_m^{(\pm)}\rangle &=& \left[
C_m^{(\pm)}\right]^2 + \left[ C_{m+1}^{(\pm)}\right]^2 = 1 \\
\langle\Psi_m^{(\mp)}\mid \Psi_m^{(\pm)}\rangle &=&
C_m^{(\pm)}C_m^{(\mp)} - C_{m+1}^{(\pm)}C_{m+1}^{(\mp)} = 0 \,. 
\end{eqnarray}
\end{mathletters}
Since $\hat {\bf S}^2$ and $\hat {\bf H}_{int}$ commute then it is
possible to find a common set of  eigenstates. We can use this fact to
determine the eigenvalues of  $\hat {\bf H}_{int}$ and the relations
among the $C$'s coefficients.  For that we need to calculate
\begin{equation}
\hat {\bf H}_{int}\mid \Psi_m^{(\pm)}\rangle =  \lambda_m^{(\pm)}\mid
\Psi_m^{(\pm)}\rangle\,,
\label{eqhipsi}
\end{equation}
where $\lambda_m^{(\pm)}$ are the eigenvalues to be determined.  Using
Eqs.~(\ref{eqso}), (\ref{eqhint}) and (\ref{eqpsm+-}), the last
eigenvalue equation can be rewritten in a matrix form as
\begin{equation}
\alpha\left[ \matrix{\beta & \hat T\hat B_- \cr \hat B_+\hat T^\dagger
& -\beta \cr}\right] \left[ \matrix{\hat T & 0\cr 0 & \pm
1\cr}\right]\left[ \matrix{C_m^{(\pm)}\mid m\rangle\cr
C_{m+1}^{(\pm)}\mid m+1\rangle}\right] =  \lambda_m^{(\pm)}\left[
\matrix{C_m^{(\pm)}\mid m\rangle\cr  C_{m+1}^{(\pm)}\mid
m+1\rangle}\right] \,,
\label{eqhimt}
\end{equation}
where $\beta = \hbar\Delta/\alpha$. Since the $C$'s coefficients
commute with the $\hat A$ or $\hat A^\dagger$ operators, then the last
matrix equation permits to obtain the following equations
\begin{mathletters}
\label{eqmat1}
\begin{eqnarray}
& &\left[ \alpha\beta-\lambda_m^{(\pm)}\right]\left(\hat T
C_m^{(\pm)}\hat T^\dagger\right)\hat T\mid m\rangle \pm
\alpha\;C_{m+1}^{(\pm)}\hat T\hat B_-\mid m+1\rangle = 0 \\ &
&\alpha\left( \hat T C_m^{(\pm)}\hat T^\dagger\right)\hat B_+\mid
m\rangle \mp \left[ \alpha\beta +
\lambda_m^{(\pm)}\right]C_{m+1}^{(\pm)} \mid m+1\rangle = 0 \,. 
\end{eqnarray}
\end{mathletters}

Introducing the operator \cite{Balantekin:1999wj} 
\begin{equation}
\hat Q^\dagger = \left(\hat B_+\hat B_-\right)^{-1/2}\hat B_+
\label{eqq+}
\end{equation}
one can write the normalized eigenstate of $\hat H_1$ as 
\begin{equation}
\mid m\rangle = \left( \hat Q^\dagger\right)^m\mid 0\rangle\,,
\label{eqsmq}
\end{equation}
and, with Eqs.~(\ref{eqq+}) and (\ref{eqsmq}) we can show that
\cite{Aleixo:2000bd} 
\begin{mathletters}
\label{eqbm}
\begin{eqnarray}
& & \hat B_+\mid m\rangle = \sqrt{{\cal E}_{m+1}}\mid m+1\rangle\,,\\
& &\hat T\hat B_-\mid m+1\rangle = \sqrt{{\cal E}_{m+1}}\,\hat T\mid
m\rangle\,. 
\end{eqnarray}
\end{mathletters}
Substituting Eqs.~(\ref{eqbm}) into Eqs.~(\ref{eqmat1}) we have 
\begin{mathletters}
\label{eqmat2}
\begin{eqnarray}
& &\left\{\left[ \alpha\beta-\lambda_m^{(\pm)}\right]\left(\hat T
C_m^{(\pm)}\hat T^\dagger\right) \pm \alpha \sqrt{{\cal
E}_{m+1}}\,C_{m+1}^{(\pm)}\right\}\hat T\mid m\rangle = 0 \\ &
&\left\{\alpha\sqrt{{\cal E}_{m+1}}\left( \hat T C_m^{(\pm)}\hat
T^\dagger\right) \mp \left[ \alpha\beta +
\lambda_m^{(\pm)}\right]C_{m+1}^{(\pm)}\right\} \mid m+1\rangle = 0
\,. 
\end{eqnarray}
\end{mathletters}
From Eqs.~(\ref{eqmat2}) it follows that 
\begin{equation}
\lambda_m^{(\pm)} = \pm\alpha\sqrt{{\cal E}_{m+1} + \beta^2}\,,
\label{eqlb}
\end{equation}
and 
\begin{equation}
C_{m+1}^{(\pm)} = \left( {\sqrt{{\cal E}_{m+1} + \beta^2}\mp
\beta\over \sqrt{{\cal E}_{m+1}}}\right)\, \left(\hat T
C_m^{(\pm)}\hat T^\dagger \right)\,. 
\label{eqccm}
\end{equation}
Eqs.~(\ref{eqcpm}) and (\ref{eqccm}) imply that 
\begin{equation}
C_{m+1}^{(\pm)} = C_m^{(\mp)}\,,
\label{eqccm1}
\end{equation}
and the eigenstates and eigenvalues of the generalized nonresonant
Jaynes-Cummings Hamiltonians can be written as 
\begin{equation}
E_m^{(\pm)} = {\cal E}_{m+1} \pm \sqrt{\alpha^2\; {\cal E}_{m+1} +
\hbar^2\Delta^2}\,,
\label{eqemjc}
\end{equation}
and 
\begin{equation}
\mid \Psi_m^{(\pm)}\rangle =  \left[ \matrix{\hat T & 0\cr 0 & \pm 1
\cr}\right]  \left[ \matrix{C_m^{(\pm)}\mid m\rangle\cr
C_m^{(\mp)}\mid m+1\rangle \cr}\right], \qquad m=0,1,2, \cdots 
\label{eqesjc}
\end{equation}
\bigskip

{\bf a) The Resonant Limit}
\bigskip

From these general results we can verify two important and simple
limiting cases. The first one corresponds to the resonant situation,
for which \ $\Delta = 0$ \ $(\beta = 0)$. Using these conditions in
Eqs.~({\ref{eqccm}) and (\ref{eqemjc}) and Eqs.~(\ref{eqcpm}) we get 
\begin{equation}
E_m^{(\pm)} = {\cal E}_{m+1} \pm \sqrt{\alpha^2\; {\cal E}_{m+1}}\,,
\label{eqemjcr}
\end{equation}
and 
\begin{equation}
C_{m+1}^{(\pm)} = \hat T C_m^{(\pm)}\hat T^\dagger = C_m^{(\pm)} =
{1\over \sqrt{2}}\,. 
\label{eqccmr}
\end{equation}
Therefore the Jaynes-Cummings resonant eigenstate is given by
\begin{equation}
\mid \Psi_m^{(\pm)}\rangle =  {1\over \sqrt{2}} \left[ \matrix{\hat T
& 0\cr 0 & \pm 1 \cr}\right]  \left[ \matrix{ \mid m\rangle\cr \mid
m+1\rangle \cr}\right], \qquad m=0,1,2, \cdots 
\label{eqesjcr}
\end{equation}
These particular results are shown in the Ref. \cite{Aleixo:2000bd}. 

\bigskip

{\bf b) The Standard Jaynes-Cummings Limit}
\bigskip

The second important limit corresponds to the standard Jaynes-Cummings
Hamiltonian, related with the harmonic oscillator system. In this
limit we have that \ $\hat T = \hat T^\dagger \longrightarrow 1$,  \
$\hat B_- \longrightarrow \hat a$, \ $\hat B_+ \longrightarrow  \hat
a^\dagger$, \ $\Delta = \omega - \omega_o$ \ and ${\cal E}_{m+1} =
(m+1)\hbar\omega$. \ Using these conditions in the
Eqs.~({\ref{eqccm}), (\ref{eqemjc}) and Eqs.~(\ref{eqcpm}) we conclude
that 
\begin{equation}
E_m^{(\pm)} = (m+1)\hbar\omega \pm \sqrt{\alpha^2 \hbar \omega (m+1) +
\hbar^2(\omega-\omega_o)^2}\,,
\label{eqemjcs}
\end{equation}
and 
\begin{equation}
C_{m+1}^{(\pm)} = \gamma_m^{(\pm)}C_m^{(\pm)} = C_m^{(\mp)} =  {1\over
\sqrt{1 + \left(\gamma_m^{(\mp)}\right)^2}}\,,
\label{eqccms}
\end{equation}
where
\begin{mathletters}
\label{eqgd}
\begin{eqnarray}
& & \gamma_m^{(\pm)} = \sqrt{1 + \delta_m^2} \mp \delta_m\,,\\ & &
\delta_m = {\hbar(\omega-\omega_o)\over \sqrt{(m+1)\alpha^2 \hbar 
\omega}}\,.
\end{eqnarray}
\end{mathletters}
Therefore the standard Jaynes-Cummings eigenstate, written in a matrix
form, is given by 
\begin{equation}
\mid \Psi_m^{(\pm)}\rangle =  {1\over \sqrt{1 +
\left(\gamma_m^{(\pm)}\right)^2}}  \left[ \matrix{ 1 & 0\cr 0 & \pm
\gamma_m^{(\pm)} \cr}\right]  \left[ \matrix{ \mid m\rangle\cr \mid
m+1\rangle \cr}\right], \qquad m=0,1,2, \cdots 
\label{eqesjcs}
\end{equation}
These results are shown in many papers, in particular, in the Ref.
\cite{ref27}. 

\section{The Time Evolution of the Nonresonant System}

To study the time-dependent Schr\"odinger equation for a
Jaynes-Cummings system in nonresonant situation  
\begin{equation}
i\hbar {\partial \over \partial t} \mid \Psi (t)\rangle =  \left(\hat
{\bf H}_o + \hat {\bf H}_{int}\right)\mid \Psi (t)\rangle
\label{eqschr}
\end{equation} 
we can write the wavefunction as 
\begin{equation}
\mid \Psi (t)\rangle = \exp{\left(-i\hat {\bf
H}_ot/\hbar\right)}\,\mid \Psi_i(t)\rangle\,,
\label{eqpsint}
\end{equation}
and, by substituting this into Schr\"odinger equation and taking into
account the commutation property between $\hat {\bf H}_o$ and $\hat
{\bf H}_{int}$, we obtain
\begin{equation}
i\hbar {\partial \over \partial t} \mid \Psi_i(t)\rangle =  \hat {\bf
H}_{int}\,\mid \Psi_i(t)\rangle\,. 
\label{eqschrint}
\end{equation} 
We introduce the evolution matrix $\hat {\bf U}_i(t,0)$: 
\begin{equation}
\mid \Psi_i(t)\rangle = \hat {\bf U}_i(t,0)\,\mid \Psi_i(0)\rangle\,. 
\label{eqevu}
\end{equation}
which satisfies the equation 
\begin{equation}
i\hbar {\partial \over \partial t} \hat {\bf U}_i(t,0) =  \hat {\bf
H}_{int}\,\hat {\bf U}_i(t,0)\,,
\label{eqschri}
\end{equation} 
that is, in matrix form, written as 
\begin{equation}
i\hbar\left[ \matrix{\hat U_{11}^\prime & \hat U_{12}^\prime \cr  \hat
U_{21}^\prime & \hat U_{22}^\prime \cr}\right] =  \alpha\left[
\matrix{\beta & \hat T\hat B_- \cr \hat B_+\hat T^\dagger & -\beta
\cr}\right]  \left[ \matrix{\hat U_{11} & \hat U_{12} \cr \hat U_{21}
& \hat U_{22} \cr}\right]\,,
\label{eqdevo}
\end{equation}
where the primes denote the time derivative.  One fast way to
diagonalize the evolution matrix differential equation is by
differentiating Eq.~({\ref{eqschri}) with respect to time. We find 
\begin{equation}
i\hbar {\partial^2 \over \partial t^2} \hat {\bf U}_i(t,0) =  \hat
{\bf H}_{int}\, {\partial \over \partial t}\hat {\bf U}_i(t,0) =
{1\over i\hbar}\hat {\bf H}_{int}^2 \hat {\bf U}_i(t,0)\,,
\label{eqschr3}
\end{equation} 
which can be written as
\begin{equation}
\left[ \matrix{\hat U_{11}^{\prime\prime} & \hat U_{12}^{\prime\prime}
\cr \hat U_{21}^{\prime\prime} & \hat U_{22}^{\prime\prime}
\cr}\right] =  -\left[ \matrix{\hat \omega_1 & 0 \cr  0 & \hat
\omega_2 \cr}\right]  \left[ \matrix{\hat U_{11} & \hat U_{12} \cr
\hat U_{21} & \hat U_{22} \cr}\right]\,,
\label{eqdevo1}
\end{equation}
where 
\begin{mathletters}
\label{eqfreq}
\begin{eqnarray}
& & \hbar\hat \omega_1 = \alpha\sqrt{\hat T\hat B_-\hat B_+\hat
T^\dagger  + \beta^2} = \sqrt{\alpha^2\,\hat H_2 +
(\hbar\Delta)^2}\,,\\ & & \hbar\hat \omega_2 = \alpha\sqrt{\hat
B_+\hat B_- + \beta^2} = \sqrt{\alpha^2\,\hat H_1 +
(\hbar\Delta)^2}\,. 
\end{eqnarray}
\end{mathletters}
Now, since by initial conditions \  $\hat {\bf U}_i(0,0) = \hat {\bf
I}$, \ then we can write the solution of the evolution matrix
differential equation ({\ref{eqschr3}) as
\begin{equation}
\hat {\bf U}_i(t,0) = \left[ \matrix{\cos{(\hat\omega_1 t)} &
\sin{(\hat\omega_1 t)}\,\hat C\cr \sin{(\hat\omega_2 t)}\,\hat D &
\cos{(\hat\omega_2 t)} \cr}\right]\,,
\label{eqevo3}
\end{equation}
and the $\hat C$ and $\hat D$ operators can be determined by the
unitarity conditions 
\begin{equation}
\hat {\bf U}_i^\dagger (t,0)\,\hat {\bf U}_i (t,0) =  \hat {\bf U}_i
(t,0)\,\hat {\bf U}_i^\dagger (t,0) = \hat {\bf I}\,. 
\label{equni}
\end{equation}
In the appendix A we show that the unitarity conditions (\ref{equni})
imply
\begin{mathletters}
\label{eqcdfin}
\begin{eqnarray}
& & \hat C = -\hat D^\dagger = {i\over {(\hat H_2)}^{1/4}}\;
\sqrt{\hat T \hat B_-}\\ & & \hat D = -\hat C^\dagger \,. 
\end{eqnarray}
\end{mathletters}
Therefore, we can write the final expression of the time evolution
matrix of the system as 
\begin{equation}
\hat {\bf U}_i(t,0) = \left[ \matrix{\cos{(\hat\omega_1 t)} &
\sin{(\hat\omega_1 t)}\,\hat C\cr -\sin{(\hat\omega_2 t)}\,\hat
C^\dagger &  \cos{(\hat\omega_2 t)} \cr}\right]\,. 
\label{eqevo4}
\end{equation}

For Jaynes-Cummings systems an important physical quantity to see how
the system under consideration evolves in time is the population
inversion factor \cite{ref21,Chaichian:1990ic,ref25}, defined by
\begin{equation}
\hat {\bf W}(t) \equiv \hat\sigma_+(t)\;\hat\sigma_-(t) -
\hat\sigma_-(t)\;\hat\sigma_+(t) = \hat\sigma_3(t)\,,
\label{eqinpo}
\end{equation}
where the time dependence of the operators is related with the
Heisenberg picture. In this case, the time evolution of the population
inversion factor will be given by 
\begin{equation}
{d\hat \sigma_3(t)\over dt} = {1\over i\hbar}\hat {\bf U}_i^\dagger
(t,0)\left[\hat\sigma_3, \hat{\bf H}\right]\hat {\bf U}_i(t,0)\,,
\label{eqds3}
\end{equation}
and since we have 
\begin{equation}
\left[ \hat\sigma_3, \hat {\bf H}\right] = \alpha\left[ \hat\sigma_3,
\hat {\bf S}\right] = -2\alpha\,\hat {\bf S}\,\hat\sigma_3\,,
\label{eqcs3h}
\end{equation}
then Eq.~(\ref{eqds3}) can be written as  
\begin{equation}
{d\hat \sigma_3(t)\over dt} = {2i\alpha\over\hbar}\;\hat {\bf S}(t)\,
\hat \sigma_3(t)\,. 
\label{eqds4}
\end{equation}
We can obtain a differential equation with constant coefficients for
$\hat\sigma_3(t)$ by taking the time derivative of Eq.~(\ref{eqds4})
\begin{equation}
{d^2\hat \sigma_3(t)\over dt^2} = {2i\alpha\over\hbar}\; \left\{
{d\hat {\bf S}(t)\over dt}\,\hat \sigma_3(t) + \hat {\bf S}(t)\,
{d\hat \sigma_3(t)\over dt}\right\}\,. 
\label{eqds5}
\end{equation}
Having in mind that 
\begin{equation}
{d\hat {\bf S}(t)\over dt} = {1\over i\hbar}\hat {\bf U}_i^\dagger
(t,0)\left[\hat {\bf S}, \hat{\bf H}\right]\hat {\bf U}_i(t,0)\,,
\label{eqds}
\end{equation}
and that 
\begin{equation}
\left[ \hat {\bf S}, \hat {\bf H}\right] = \alpha\beta\left[ \hat {\bf
S}, \hat\sigma_3\right] = 2\alpha\beta\,\hat {\bf S}\,\hat\sigma_3\,,
\label{eqcshh}
\end{equation}
we conclude that 
\begin{equation}
{d\hat {\bf S}(t)\over dt} = -{2i\alpha\beta\over\hbar}\;\hat {\bf
S}(t)\, \hat \sigma_3(t)\,. 
\label{eqdss}
\end{equation}
Using Eqs.~(\ref{eqds4}) and (\ref{eqdss}) into Eq.~(\ref{eqds5}) we
obtain 
\begin{equation}
{d^2\hat \sigma_3(t)\over dt^2} + \hat {\bf \Theta}^2\, \hat
\sigma_3(t) = \hat {\bf F}(t) 
\label{eqds6}
\end{equation}
where 
\begin{mathletters}
\label{eqomft}
\begin{eqnarray}
& & \hat {\bf \Theta}^2 = {4\alpha^2\over\hbar^2}\,\hat {\bf S}^2\\ &
& \hat {\bf F}(t) = {4\alpha^2\beta\over\hbar^2}\, \hat {\bf
U}_i^\dagger (t,0)\,\hat {\bf S}\,\hat {\bf U}_i(t,0)\,. 
\end{eqnarray}
\end{mathletters}

Eq.~(\ref{eqds6}) corresponds to a non-homogeneous linear differential
equation for $\hat \sigma_3(t)$ with constant coefficients since $\hat
{\bf S}^2$ and $\hat {\bf H}$ commute and, therefore, $\hat {\bf
\Theta }$ is a constant of the motion. The general solution of this
differential equation can be written as 
\begin{equation}
\hat \sigma_3(t) = \hat\sigma^H(t) + \hat\sigma^P(t)\,,
\label{eqs3gen}
\end{equation}
and each matrix element of the homogeneous solution, satisfies the
differential equation 
\begin{equation}
{d^2\hat\sigma^H_{jk}(t)\over dt^2} +
\hat\nu_j^2\,\hat\sigma^H_{jk}(t) = 0\,,  \qquad j,\,k = 1, {\rm or}\;
2\,,
\label{eqs3ho}
\end{equation}
with 
\begin{mathletters}
\label{eqfreqo}
\begin{eqnarray}
& & \hbar\hat \nu_1 = 2\alpha\sqrt{\hat T\hat B_-\hat B_+\hat
T^\dagger} = 2\sqrt{\alpha^2\,\hat H_2}\,,\\ & & \hbar\hat \nu_2 =
2\alpha\sqrt{\hat B_+\hat B_-} = 2\sqrt{\alpha^2\,\hat H_1}\,. 
\end{eqnarray}
\end{mathletters}
The solution of Eq.~(\ref{eqs3ho}) is given by 
\begin{equation}
\hat\sigma^H_{jk}(t) = \hat y_j(t)\;\hat c_{jk} + \hat z_j(t)\;\hat
d_{jk}\,,
\label{eqs3hog}
\end{equation}
where  
\begin{mathletters}
\label{eqlisol}
\begin{eqnarray}
& & \hat y_j(t) = \cos{(\hat\nu_jt)}\\ & & \hat z_j(t) =
\sin{(\hat\nu_jt)}\,,
\end{eqnarray}
\end{mathletters}
and the coefficients $\hat c_{jk}$ and $\hat d_{jk}$ can be determined
by the initial conditions. 

The matrix elements of the particular solution of the
$\hat\sigma_3(t)$ differential equation need to satisfy  
\begin{equation}
{d^2\hat\sigma^P_{jk}(t)\over dt^2} +
\hat\nu_j^2\,\hat\sigma^P_{jk}(t) = \hat F_{jk}(t)\,,  \qquad j,\,k =
1, {\rm or}\; 2\,,
\label{eqs3pt}
\end{equation}
and they can be obtained by the variation of parameter or by Green
function methods, giving
\begin{equation}
\hat\sigma_{jk}^P(t) = \hat\nu_j^{-1}\,\left\{ \hat z_j(t) \int_0^t
d\xi\,\hat y_j(\xi)\,\hat F_{jk}(\xi) -  \hat y_j(t)\int_0^t
d\xi\,\hat z_j(\xi)\,\hat F_{jk}(\xi) \right\}\,,
\label{eqvarp}
\end{equation}
where we used that the Wronskian of the system of solutions $\hat
y_j(t)$ and $\hat z_j(t)$ is given by $\hat\nu_j$. 

After we determine the elements of the $\hat {\bf F}(t)$-matrix, it is
necessary to resolve the integrals in Eq.~(\ref{eqvarp}) to obtain the
explicit expression of the particular solution.  In the appendix B we
show that, using Eqs.~(\ref{eqso}), (\ref{eqevo4}), and
(\ref{eqomft}), it is possible to conclude that these matrix elements
can be written as 
\begin{mathletters}
\label{eqs3p11}
\begin{eqnarray}
\hat\sigma_{11}^P(t) &=& i{\gamma\over 2}\hat\nu_1^{-1}\sqrt{\hat
T\hat B_-}\left\{\hat z_2(t)\,{\cal
G}_{CS}^{(+)}(t;\hat\nu_2,\hat\omega_2,\hat\omega_1) -  \hat
y_2(t)\,{\cal G}_{SS}^{(+)}(t;\hat\nu_2,\hat\omega_2, \hat\omega_1)
\right\}\hat H_2^{1/4}\nonumber\\ &+& i{\gamma\over
2}\hat\nu_1^{-1}\hat H_2^{1/4}\left\{\hat z_1(t)\, {\cal
G}_{SC}^{(-)}(t;\hat\nu_1,\hat\omega_1,\hat\omega_2) -  \hat
y_1(t)\,{\cal G}_{CC}^{(-)}(t;\hat\nu_1,\hat\omega_1, \hat\omega_2)
\right\}\sqrt{\hat B_+\hat T^\dagger}\,,\\
\hat\sigma_{12}^P(t) &=& {\gamma\over 2}\hat\nu_1^{-1}\sqrt{\hat T\hat
B_-}\left\{\hat z_2(t)\,{\cal
G}_{CC}^{(+)}(t;\hat\nu_2,\hat\omega_2,\hat\omega_1) -  \hat
y_2(t)\,{\cal G}_{SC}^{(+)}(t;\hat\nu_2,\hat\omega_2, \hat\omega_1)
\right\}\sqrt{\hat T\hat B_-}\nonumber\\ &+& {\gamma\over
2}\hat\nu_1^{-1}\hat H_2^{1/4}\left\{\hat z_1(t)\, {\cal
G}_{SS}^{(-)}(t;\hat\nu_1,\hat\omega_1,\hat\omega_2) +  \hat
y_1(t)\,{\cal G}_{CS}^{(-)}(t;\hat\nu_1,\hat\omega_1, \hat\omega_2)
\right\}\hat H_1^{1/4}\,,\\
\hat\sigma_{21}^P(t) &=& {\gamma\over 2}\hat\nu_2^{-1}\sqrt{\hat
B_+\hat T^\dagger}\left\{\hat z_1(t)\,{\cal
G}_{CC}^{(+)}(t;\hat\nu_1,\hat\omega_1,\hat\omega_2) -  \hat
y_1(t)\,{\cal G}_{SC}^{(+)}(t;\hat\nu_1,\hat\omega_1, \hat\omega_2)
\right\}\sqrt{\hat B_+\hat T^\dagger}\nonumber\\ &+& {\gamma\over
2}\hat\nu_2^{-1}\hat H_1^{1/4}\left\{\hat z_2(t)\, {\cal
G}_{SS}^{(-)}(t;\hat\nu_2,\hat\omega_2,\hat\omega_1) -  \hat
y_2(t)\,{\cal G}_{CS}^{(-)}(t;\hat\nu_2,\hat\omega_2, \hat\omega_1)
\right\}\hat H_2^{1/4}\,,\\
\hat\sigma_{22}^P(t) &=& i{\gamma\over 2}\hat\nu_2^{-1}\sqrt{\hat
B_+\hat T^\dagger}\left\{\hat z_1(t)\,{\cal
G}_{CS}^{(+)}(t;\hat\nu_1,\hat\omega_1,\hat\omega_2) -  \hat
y_1(t)\,{\cal G}_{SS}^{(+)}(t;\hat\nu_1,\hat\omega_1, \hat\omega_2)
\right\}\hat H_1^{1/4}\nonumber\\ &+& i{\gamma\over
2}\hat\nu_2^{-1}\hat H_1^{1/4}\left\{\hat z_2(t)\, {\cal
G}_{SC}^{(-)}(t;\hat\nu_2,\hat\omega_2,\hat\omega_1) +  \hat
y_2(t)\,{\cal G}_{CC}^{(-)}(t;\hat\nu_2,\hat\omega_2, \hat\omega_1)
\right\}\sqrt{\hat T\hat B_-}\,,
\end{eqnarray}
\end{mathletters}
where \ $\gamma = 4\alpha^2\beta/\hbar^2$, \  and the auxiliary
functions are given by 
\begin{equation}
{\cal G}^{(\pm)}_{XY}(t;\hat p,\hat q,\hat r) = {\cal F}_{XY}(t;\hat
p-\hat q,\hat r) \pm {\cal F}_{XY}(t;\hat p+\hat q,\hat r)\,,\qquad X,
Y = C\;{\rm or}\;S\,,
\label{eqfgpm}
\end{equation}
with 
\begin{mathletters}
\label{eqfxy}
\begin{eqnarray}
{\cal F}_{CC}(t;\hat x,\hat w) &\equiv& \int_0^td\xi\,\cos{(\hat
x\xi)}\, \cos{(\hat w\xi)}\nonumber\\ &=& \sum_{m,n = 0}^\infty
(-1)^{m+n}{\hat x^{2m}\hat  w^{2n}\over (2m)!\,(2n)!}{t^{2m+2n+1}\over
(2m+2n+1)}\,\\
{\cal F}_{CS}(t;\hat x,\hat w) &\equiv& \int_0^td\xi\,\cos{(\hat
x\xi)}\, \sin{(\hat w\xi)}\nonumber\\ &=& \sum_{m,n = 0}^\infty
(-1)^{m+n}{\hat x^{2m}\hat  w^{2n+1}\over
(2m)!\,(2n+1)!}{t^{2m+2n+2}\over (2m+2n+2)}\,\\
{\cal F}_{SC}(t;\hat x,\hat w) &\equiv& \int_0^td\xi\,\sin{(\hat
x\xi)}\, \cos{(\hat w\xi)}\nonumber\\ &=& \sum_{m,n = 0}^\infty
(-1)^{m+n}{\hat x^{2m+1}\hat  w^{2n}\over
(2m+1)!\,(2n)!}{t^{2m+2n+2}\over (2m+2n+2)}\,\\
{\cal F}_{SS}(t;\hat x,\hat w) &\equiv& \int_0^td\xi\,\sin{(\hat
x\xi)}\, \sin{(\hat w\xi)}\nonumber\\ &=& \sum_{m,n = 0}^\infty
(-1)^{m+n}{\hat x^{2m+1}\hat  w^{2n+1}\over
(2m+1)!\,(2n+1)!}{t^{2m+2n+3}\over (2m+2n+3)}\,. 
\end{eqnarray}
\end{mathletters}
With these results for the particular solution we can conclude that 
\begin{equation}
\hat\sigma^P_{ij}(0) = 0 = {d\hat\sigma^P_{ij}(0)\over dt}\,. 
\label{eqsp0}
\end{equation}
Now, using Eqs.~(\ref{eqds4}), (\ref{eqs3gen}), (\ref{eqs3hog}),
(\ref{eqsp0}) and the initial conditions, we have 
\begin{mathletters}
\label{eqs3ini}
\begin{eqnarray}
& & \left[\hat\sigma_3(0)\right]_{ij} = \hat c_{ij}\\ & &
\left[{d\hat\sigma_3(0)\over dt}\right]_{ij} = {2i\alpha\over \hbar}\,
\left[\hat {\bf S}(0)\,\hat\sigma_3(0)\right]_{ij} = \hat\nu_i\,\hat
d_{ij}\,. 
\end{eqnarray}
\end{mathletters}
Therefore, the final expression for the elements of the population
inversion matrix of the system can be written as 
\begin{equation}
[\hat\sigma_3(t)]_{ij} = \cos{(\hat\nu_it)}\,
\left[\hat\sigma_3(0)\right]_{ij} +  {2i\alpha\over
\hbar}\,\sin{(\hat\nu_it)}\; \hat\nu_i^{-1} \left[\hat {\bf
S}(0)\,\hat\sigma_3(0)\right]_{ij} + \hat\sigma^P_{ij}(t)\,. 
\label{eqs3fin}
\end{equation}

Again, using these final results we can verify two important and
simple limit cases. 

\bigskip

{\bf a) The Resonant Limit}
\bigskip

The first one corresponds to the resonant situation $(\Delta = 0)$.
Eqs.~({\ref{eqfreq}), (\ref{eqevo4}), (\ref{eqfreqo}) and
(\ref{eqs3p11}) allow us to conclude that, in this case, the evolution
matrix of the system is given by 
\begin{equation}
\hat {\bf U}_i(t,0) = \left[ \matrix{\cos{\left({1\over
2}\hat\nu_1t\right)} &  \sin{\left({1\over 2}  \hat\nu_1 t \right)}\,
\hat C\cr -\sin{\left({1\over 2}\hat\nu_2 t\right)}\,  \hat C^\dagger
& \cos{\left({1\over 2}\hat\nu_2 t\right)} \cr}\right]\,. 
\label{eqevo5}
\end{equation}
and the elements of the population inversion of the system are 
\begin{equation}
[\hat\sigma_3(t)]_{ij} = \cos{(\hat\nu_it)}\,
\left[\hat\sigma_3(0)\right]_{ij} +  {2i\alpha\over
\hbar}\,\sin{(\hat\nu_it)}\; \hat\nu_i^{-1} \left[\hat {\bf
S}(0)\,\hat\sigma_3(0)\right]_{ij}\,. 
\label{eqs3fres}
\end{equation}

\bigskip

{\bf b) The Standard Jaynes-Cummings Limit}
\bigskip

This second important limit corresponds to the case of the harmonic
oscillator system, and in this limit we have that  \ $\hat T = \hat
T^\dagger \longrightarrow 1$, \ $\hat B_- \longrightarrow \hat a$, \
$\hat B_+ \longrightarrow \hat a^\dagger$ \ and \ $[\hat a,\hat
a^\dagger] = \hbar\omega$. \  With these conditions the operators
$\hat\omega_1$  and $\hat\omega_2$ commute, and this fact permits to
evaluate the integrals related with the particular solution of the
population inversion elements using trigonometric product
relations. Using that and the expressions obtained in the appendix B,
after a considerable amount of algebra and trigonometric product
relations we can show that is possible to write the expressions for
the $\hat\sigma_{ij}^P(t)$-matrix elements as 
\begin{mathletters}
\label{eqs3p11s}
\begin{eqnarray}
\hat\sigma_{11}^P(t) &=& i{\gamma\over 2}\hat\nu_1^{-1} \sqrt{\hat
a}\left\{{\cal K}_S(t;\hat\omega_2,\hat\omega_1,\hat\nu_2) - {\cal
K}_S(t;\hat\omega_2,-\hat\omega_1,\hat\nu_2) \right\} \left(\hat a\hat
a^\dagger\right)^{1/4}\nonumber\\ &-& i{\gamma\over
2}\hat\nu_1^{-1}\left(\hat a\hat a^\dagger\right)^{1/4} \left\{{\cal
K}_S(t;\hat\omega_2,\hat\omega_1,\hat\nu_1) - {\cal
K}_S(t;\hat\omega_2,-\hat\omega_1,\hat\nu_1) \right\} \sqrt{\hat
a^\dagger }\\
\hat\sigma_{12}^P(t) &=& {\gamma\over 2}\hat\nu_1^{-1}\sqrt{\hat a}
\left\{{\cal K}_C(t;\hat\omega_2,\hat\omega_1,\hat\nu_2) - {\cal
K}_C(t;\hat\omega_2,-\hat\omega_1,\hat\nu_2)\right\} \sqrt{\hat
a}\nonumber\\ &-& {\gamma\over 2}\hat\nu_1^{-1}\left(\hat a\hat
a^\dagger\right)^{1/4} \left\{{\cal
K}_C(t;\hat\omega_2,\hat\omega_1,\hat\nu_1) - {\cal
K}_C(t;\hat\omega_2,-\hat\omega_1,\hat\nu_1)\right\} \left(\hat
a^\dagger\hat a\right)^{1/4}\\
\hat\sigma_{21}^P(t) &=& {\gamma\over 2}\hat\nu_2^{-1}\sqrt{\hat
a^\dagger} \left\{{\cal K}_C(t;\hat\omega_2,\hat\omega_1,\hat\nu_1) +
{\cal K}_C(t;\hat\omega_2,-\hat\omega_1,\hat\nu_1)\right\} \sqrt{\hat
a^\dagger}\nonumber\\ &-& {\gamma\over 2}\hat\nu_2^{-1}\left(\hat
a^\dagger\hat a\right)^{1/4} \left\{{\cal
K}_C(t;\hat\omega_2,\hat\omega_1,\hat\nu_2) - {\cal
K}_C(t;\hat\omega_2,-\hat\omega_1,\hat\nu_2)\right\} \left(\hat a\hat
a^\dagger\right)^{1/4}\\
\hat\sigma_{22}^P(t) &=& i{\gamma\over 2}\hat\nu_2^{-1} \sqrt{\hat
a^\dagger}\left\{{\cal K}_S(t;\hat\omega_2,\hat\omega_1,\hat\nu_1)  +
{\cal K}_S(t;\hat\omega_2,-\hat\omega_1,\hat\nu_1)\right\} \left(\hat
a^\dagger\hat a\right)^{1/4}\nonumber\\ &-& i{\gamma\over
2}\hat\nu_2^{-1}\left(\hat a^\dagger\hat a\right)^{1/4} \left\{{\cal
K}_S(t;\hat\omega_2,\hat\omega_1,\hat\nu_2) + {\cal
K}_S(t;\hat\omega_2,-\hat\omega_1,\hat\nu_2) \right\}\sqrt{\hat a}\,,
\end{eqnarray}
\end{mathletters}
where, now, the auxiliary functions are given by 
\begin{mathletters}
\label{eqksc}
\begin{eqnarray}
{\cal K}_S(t;\hat p,\hat q,\hat r) &=& {\hat r\;\sin{\left [\left(\hat
p+\hat q\right)t\right ]}-\left(\hat p+\hat q\right)\,\sin{\left(\hat
rt\right)}\over \hat r^2-\left(\hat p+\hat q\right)^2}\\
{\cal K}_C(t;\hat p,\hat q,\hat r) &=& {\hat r\;\cos{\left [\left(\hat
p+\hat q\right)t\right ]}-\hat r\;\cos{\left(\hat rt\right)}\over \hat
r^2-\left(\hat p+\hat q\right)^2}\,. 
\end{eqnarray}
\end{mathletters}
Considering the expressions above we may easily verify that the
particular solution for the population inversion factor must still
satisfy the initial conditions (\ref{eqsp0}). Therefore, in this case
the final expression for the population inversion factor has the same
form given by Eq.~(\ref{eqs3fin}), with  
\begin{mathletters}
\label{eqfrosc}
\begin{eqnarray}
& & \hbar\hat \nu_1 = 2\alpha\sqrt{\hat a\hat a^\dagger
}\,,\qquad\qquad\qquad \hbar\hat \nu_2 = 2\alpha\sqrt{\hat
a^\dagger\hat a}\,,\\ & & \hbar\hat \omega_1 = \alpha\sqrt{\hat a\hat
a^\dagger  + \beta^2}\,,\qquad\qquad  \hbar\hat \omega_2 =
\alpha\sqrt{\hat a^\dagger\hat a +\beta^2}\,. 
\end{eqnarray}
\end{mathletters}

\section{The Generalized Intensity-Dependent Nonresonant 
Jaynes-Cummings Hamiltonian}

A variant of the Jaynes-Cummings model takes the coupling between
matter and the radiation to depend on the intensity of the
electromagnetic field \cite{Chaichian:1990ic,ref25,ref26,ref30}. This
model has 
great relevance since this kind of interaction means effectively that
the coupling is proportional to the amplitude of the field which is a
very simple case of a nonlinear interaction corresponding to a more
realistic physical situation. The results of this model can also give
insight into the behavior of other quantum systems with strong
nonlinear interactions. In this section we generalize the standard
intensity-dependent nonresonant Jaynes-Cummings model to a
shape-invariant one.

The expression for the intensity-dependent nonresonant
Jaynes-Cummings Hamiltonian can be written as
\begin{equation}
\hat {\bf H} = \hat A^\dagger\hat A + {1\over 2}\left[\hat A,\hat
A^\dagger\right]\left(\hat\sigma_3+1\right) + \alpha\left(\hat\sigma_+\hat
A\;\sqrt{\hat A^\dagger\hat A}\, + \hat\sigma_-\sqrt{\hat A^\dagger\hat
A}\;\hat A^\dagger\right) + \hbar\Delta\,\hat\sigma_3\,.
\label{eqjcnrid}
\end{equation}
Note that here the constant $\alpha$ is dimensionless. To generalize
Hamiltonian (\ref{eqjcnrid}) for all supersymmetric and
shape-invariant systems, we can use the operator $\hat {\bf S}$, given
by Eq.~(\ref{eqso}), and further by introducing the following operator

\begin{equation}
\hat {\bf S}_i = \hat\sigma_+\hat A\;\sqrt{\hat A^\dagger\hat A}\, +
\hat\sigma_-\sqrt{\hat A^\dagger\hat A}\;\hat A^\dagger\,.
\label{eqSi}
\end{equation}
Again, the operators $\hat A$ and $\hat A^\dagger$ satisfy the shape
invariance condition, Eq.~(\ref{eqsi}). Using operators $\hat {\bf S}$
and $\hat {\bf S}_i$ we can decompose the Jaynes-Cummings Hamiltonian
in the form

\begin{equation}
\hat {\bf H} = \hat {\bf H}_o + \hat {\bf H}_{int}\,,
\label{eqjcnrid1}
\end{equation}
where 
\begin{mathletters}
\label{eqhintid}
\begin{eqnarray}
& & \hat {\bf H}_o = \hat {\bf S}^2\,,\\ & & \hat {\bf H}_{int} =
\alpha\,\hat {\bf S}_i + \hbar\Delta\,\hat\sigma_3\,. 
\end{eqnarray}
\end{mathletters}
In this case, $\hat {\bf H}_{int}$ can be written as

\begin{equation}
\hat {\bf H}_{int} = \alpha \left[ \matrix{\beta & \hat T\hat
B_-\;\sqrt{\hat B_+\hat B_-}\cr \sqrt{\hat B_+\hat B_-}\;\hat B_+ \hat
T^\dagger & -\beta\cr}\right]\; = \;\alpha \left[ \matrix{\beta & \hat
T\hat B_-\;\sqrt{\hat H_1}\cr \sqrt{\hat H_1}\;\hat B_+ \hat T^\dagger
& -\beta\cr}\right]\,.
\label{eqHintid}
\end{equation}
Here we can follow the same development of the section II, with the
same notation. So, using Eqs.~(\ref{eqso}), (\ref{eqpsm+-}),
(\ref{eqSi}) and (\ref{eqhintid}), the eigenvalue equation

\begin{equation}
\hat {\bf H}_{int}\mid \Psi_m^{(\pm)}\rangle =  \lambda_m^{(\pm)}\mid
\Psi_m^{(\pm)}\rangle\,,
\label{eqhipsiid}
\end{equation}
can be written in a matrix form as

\begin{equation}
\alpha\left[ \matrix{\beta & \hat T\hat B_-\;\sqrt{\hat H_1} \cr
\sqrt{\hat H_1}\;\hat B_+\hat T^\dagger & -\beta \cr}\right] \left[
\matrix{\hat T & 0\cr 0 & \pm 1\cr}\right]\left[
\matrix{C_m^{(\pm)}\mid m\rangle\cr  C_{m+1}^{(\pm)}\mid
m+1\rangle}\right] =  \lambda_m^{(\pm)}\left[ \matrix{C_m^{(\pm)}\mid
m\rangle\cr  C_{m+1}^{(\pm)}\mid m+1\rangle}\right] \,.
\label{eqhimtid}
\end{equation}
Again, since the $C$'s coefficients commute with the $\hat A$ or $\hat
A^\dagger$ operators, then the last matrix equation permits to obtain
the following equations

\begin{mathletters}
\label{eqmat1id}
\begin{eqnarray}
& &\left[ \alpha\beta-\lambda_m^{(\pm)}\right]\left(\hat T
C_m^{(\pm)}\hat T^\dagger\right)\hat T\mid m\rangle \pm
\alpha\,C_{m+1}^{(\pm)}\hat T\hat B_-\;\sqrt{\hat H_1} \mid m+1\rangle= 0 \\ & &\alpha\left( \hat T C_m^{(\pm)}\hat T^\dagger\right)
\sqrt{\hat H_1}\;\hat B_+\mid m\rangle \mp \left[ \alpha\beta +
\lambda_m^{(\pm)}\right]C_{m+1}^{(\pm)} \mid m+1\rangle = 0 \,. 
\end{eqnarray}
\end{mathletters}

Now, using from Eqs.~(\ref{eqq+}) to (\ref{eqbm}) we have 

\begin{eqnarray}
\hat T\hat B_-\;\sqrt{\hat H_1}\mid m+1\rangle &=&   \hat T\hat
B_-\;\sqrt{{\cal E}_{m+1}}\mid m+1\rangle\nonumber\\ &=& \sqrt{{\cal
E}_{m+1}}\;\hat T\hat B_-\mid m+1\rangle\nonumber\\ &=& {\cal
E}_{m+1}\;\hat T\mid m\rangle\,,
\label{eqtbh1id}
\end{eqnarray}
and 
\begin{eqnarray}
\sqrt{\hat H_1}\;\hat B_+\mid m\rangle &=&   \sqrt{\hat
H_1}\;\sqrt{{\cal E}_{m+1}}\mid m+1\rangle\nonumber\\ &=& \sqrt{{\cal
E}_{m+1}}\;\sqrt{\hat H_1}\mid m+1\rangle\nonumber\\ &=& {\cal
E}_{m+1}\mid m+1\rangle\,,
\label{eqh1b+id}
\end{eqnarray}
Using Eqs.~(\ref{eqtbh1id}) and (\ref{eqh1b+id}), then
Eqs.~(\ref{eqmat1id}) take the form

\begin{mathletters}
\label{eqmat2id}
\begin{eqnarray}
& &\left\{\left[ \alpha\beta-\lambda_m^{(\pm)}\right]\left(\hat T
C_m^{(\pm)}\hat T^\dagger\right) \pm \alpha \;{\cal
E}_{m+1}\,C_{m+1}^{(\pm)}\right\}\hat T\mid m\rangle = 0 \\ &
&\left\{\alpha \;{\cal E}_{m+1}\left( \hat T C_m^{(\pm)}\hat
T^\dagger\right) \mp \left[ \alpha\beta +
\lambda_m^{(\pm)}\right]C_{m+1}^{(\pm)}\right\} \mid m+1\rangle = 0
\,. 
\end{eqnarray}
\end{mathletters}
From Eqs.~(\ref{eqmat2id}) it follows that

\begin{equation}
\lambda_m^{(\pm)} = \pm\alpha\sqrt{{\cal E}_{m+1}^2 + \beta^2}\,,
\label{eqlbid}
\end{equation}
and 
\begin{equation}
C_{m+1}^{(\pm)} = \left( {\sqrt{{\cal E}_{m+1}^2 + \beta^2}\mp
\beta\over {\cal E}_{m+1}}\right)\, \left(\hat T C_m^{(\pm)}\hat
T^\dagger \right)\,. 
\label{eqccmid}
\end{equation}
Eqs.~(\ref{eqcpm}) and (\ref{eqccmid}) imply that

\begin{equation}
C_{m+1}^{(\pm)} = C_m^{(\mp)}\,,
\label{eqccm1id}
\end{equation}
and the eigenstates and eigenvalues of the generalized
intensity-dependent nonresonant Jaynes-Cummings Hamiltonian can be
written as

\begin{equation}
E_m^{(\pm)} = {\cal E}_{m+1} \pm \sqrt{\alpha^2\; {\cal E}_{m+1}^2 +
\hbar^2\Delta^2}\,,
\label{eqemjcid}
\end{equation}
and 
\begin{equation}
\mid \Psi_m^{(\pm)}\rangle =  \left[ \matrix{\hat T & 0\cr 0 & \pm 1
\cr}\right]  \left[ \matrix{C_m^{(\pm)}\mid m\rangle\cr
C_m^{(\mp)}\mid m+1\rangle \cr}\right], \qquad m=0,1,2, \cdots 
\label{eqesjcid}
\end{equation}

\bigskip
{\bf a) The Intensity-Dependent Resonant Limit}
\bigskip

From these general results we can again verify the two simple limiting
cases. The first one, corresponding to the resonant situation, is
for \ $\Delta = 0$ \ $(\beta = 0)$. Using these conditions into
Eqs.~({\ref{eqccmid}) and (\ref{eqemjcid}) and Eqs.~(\ref{eqcpm}) we can
promptly conclude that

\begin{equation}
E_m^{(\pm)} = \left( 1 \pm \alpha\right)\,{\cal E}_{m+1}\,,
\label{eqemjcrid}
\end{equation}
and 
\begin{equation}
C_{m+1}^{(\pm)} = \hat T C_m^{(\pm)}\hat T^\dagger = C_m^{(\pm)} =
{1\over \sqrt{2}}\,. 
\label{eqccmrid}
\end{equation}
Therefore the intensity-dependent resonant Jaynes-Cummings eigenstate
is given by
\begin{equation}
\mid \Psi_m^{(\pm)}\rangle =  {1\over \sqrt{2}} \left[ \matrix{\hat T
& 0\cr 0 & \pm 1 \cr}\right]  \left[ \matrix{ \mid m\rangle\cr \mid
m+1\rangle \cr}\right], \qquad m=0,1,2, \cdots \cdots\,.
\label{eqesjcrid}
\end{equation}
If we compare this last particular result with that one found in the
reference \cite{Aleixo:2000bd}, we conclude that the
intensity-dependent and 
intensity-independent generalized Jaynes-Cummings Hamiltonians have
the same eigenstates in the resonant situation.

\bigskip

{\bf b) The Standard Intensity-Dependent Jaynes-Cummings Limit}
\bigskip

The second limit, corresponding to the standard
intensity-dependent Jaynes-Cummings case, is related with the harmonic
oscillator system. In this limit we have that  \ $\hat T = \hat
T^\dagger \longrightarrow 1$, \ $\hat B_- \longrightarrow \hat a$, \
$\hat B_+ \longrightarrow \hat a^\dagger$, \ $\Delta = \omega -
\omega_o$ \ and ${\cal E}_{m+1} = (m+1)\hbar\omega$. \ Using these
conditions in  Eqs.~({\ref{eqccmid}), (\ref{eqemjcid}) and
Eqs.~(\ref{eqcpm}) we  can promptly conclude that

\begin{equation}
E_m^{(\pm)} = (m+1)\hbar\omega \pm \hbar\sqrt{\alpha^2\omega^2 (m+1)^2
+  (\omega-\omega_o)^2}\,,
\label{eqemjcsid}
\end{equation}
and  
\begin{equation}
C_{m+1}^{(\pm)} = \gamma_m^{(\pm)}C_m^{(\pm)} = C_m^{(\mp)} =  {1\over
\sqrt{1 + \left(\gamma_m^{(\mp)}\right)^2}}\,,
\label{eqccmsid}
\end{equation}
where 
\begin{mathletters}
\label{eqgdid}
\begin{eqnarray}
& & \gamma_m^{(\pm)} = \sqrt{1 + \delta_m^2} \mp \delta_m\,,\\ & &
\delta_m = {\omega-\omega_o\over \alpha\omega (m+1)}\,. 
\end{eqnarray}
\end{mathletters}
Therefore the standard intensity-dependent Jaynes-Cummings eigenstate,
written in a matrix form, is given by 
\begin{equation}
\mid \Psi_m^{(\pm)}\rangle =  {1\over \sqrt{1 +
\left(\gamma_m^{(\pm)}\right)^2}}  \left[ \matrix{ 1 & 0\cr 0 & \pm
\gamma_m^{(\pm)} \cr}\right]  \left[ \matrix{ \mid m\rangle\cr \mid
m+1\rangle \cr}\right], \qquad m=0,1,2, \cdots\,. 
\label{eqesjcsid}
\end{equation}

\section{Time Evolution of the Intensity-Dependent Nonresonant System}

To resolve the time-dependent Schr\"odinger equation for
intensity-dependent nonresonant Jaynes-Cummings systems:  
\begin{equation}
i\hbar {\partial \over \partial t} \mid \Psi (t)\rangle =  \left(\hat
{\bf H}_o + \hat {\bf H}_{int}\right)\mid \Psi (t)\rangle
\label{eqschrid}
\end{equation} 
we can again write the state $\mid \Psi (t)\rangle$ as it is given by
Eq.~(\ref{eqpsint}). Then using from Eqs.~(\ref{eqschrint}) to
(\ref{eqschri}), we can write the matrix equation

\begin{equation}
i\hbar\left[ \matrix{\hat U_{11}^\prime & \hat U_{12}^\prime \cr  \hat
U_{21}^\prime & \hat U_{22}^\prime \cr}\right] =  \alpha\left[
\matrix{\beta & \hat T\hat B_-\;\sqrt{\hat H_1} \cr \sqrt{\hat
H_1}\;\hat B_+\hat T^\dagger & -\beta \cr}\right]  \left[ \matrix{\hat
U_{11} & \hat U_{12} \cr \hat U_{21} & \hat U_{22} \cr}\right]\,,
\label{eqdevoid}
\end{equation}
To diagonalize this evolution matrix differential equation
we can differentiate Eq.~({\ref{eqschri}) with respect to time. After
that, if we again use the same Eq.~({\ref{eqschri}), we find

\begin{equation}
i\hbar {\partial^2 \over \partial t^2} \hat {\bf U}_i(t,0) =  \hat
{\bf H}_{int}\, {\partial \over \partial t}\hat {\bf U}_i(t,0) =
{1\over i\hbar}\hat {\bf H}_{int}^2 \hat {\bf U}_i(t,0)\,,
\label{eqschr3id}
\end{equation}
which can be written as 
\begin{equation}
\left[ \matrix{\hat U_{11}^{\prime\prime} & \hat U_{12}^{\prime\prime}
\cr \hat U_{21}^{\prime\prime} & \hat U_{22}^{\prime\prime}
\cr}\right] =  -\left[ \matrix{\hat \omega_1 & 0 \cr  0 & \hat
\omega_2 \cr}\right]  \left[ \matrix{\hat U_{11} & \hat U_{12} \cr
\hat U_{21} & \hat U_{22} \cr}\right]\,,
\label{eqdevo1id}
\end{equation}
where 
\begin{mathletters}
\label{eqfreqid}
\begin{eqnarray}
& & \hbar\hat \omega_1 = \alpha\sqrt{(\hat T\hat B_-\hat B_+\hat
T^\dagger)^2  + \beta^2} = \sqrt{\alpha^2\,\hat H_2^2 +
(\hbar\Delta)^2}\,,\\ & & \hbar\hat \omega_2 = \alpha\sqrt{(\hat
B_+\hat B_-)^2 + \beta^2} = \sqrt{\alpha^2\,\hat H_1^2 +
(\hbar\Delta)^2}\,. 
\end{eqnarray}
\end{mathletters}
Now, using the initial conditions $\hat {\bf U}_i(0,0) = \hat {\bf
I}$, we can write the solution of the evolution matrix differential
equation ({\ref{eqschr3id}) as

\begin{equation}
\hat {\bf U}_i(t,0) = \left[ \matrix{\cos{(\hat\omega_1 t)} &
\sin{(\hat\omega_1 t)}\,\hat C\cr \sin{(\hat\omega_2 t)}\,\hat D &
\cos{(\hat\omega_2 t)} \cr}\right]\,,
\label{eqevo3id}
\end{equation}
where the $\hat C$ and $\hat D$ operators can be determined by
Eq.~(\ref{equni}). Following the same steps used in the appendix A, we
can conclude that these operators must have the form given by
Eqs.~(\ref{eqcdfin}). So, in this case the final expression of the
time evolution matrix $\hat {\bf U}_i(t,0)$ is given by
Eq.~(\ref{eqevo4}) as well.

To obtain the population inversion factor we can again follow the
steps from Eq.~(\ref{eqinpo}) to (\ref{eqvarp}), but replacing the
operator $\hat {\bf S}$ by the operator $\hat {\bf S}_i$. Besides that
we have

\begin{mathletters}
\label{eqfreqoid}
\begin{eqnarray}
& & \hbar\hat \nu_1 = 2\alpha\,\hat T\hat B_-\hat B_+\hat T^\dagger =
2\alpha\,\hat H_2\,,\\ & & \hbar\hat \nu_2 = 2\alpha\,\hat B_+\hat B_-
= 2\alpha\,\hat H_1\,, 
\end{eqnarray}
\end{mathletters}
instead Eqs.~(\ref{eqfreqo}). Here, we can again use the development
shown in the appendix B, just replacing $\hat {\bf S}$ by $\hat {\bf
S}_i$, to obtain the explicit form of the matrix elements for the
particular solution of the population inversion factor, given by
Eq.~(\ref{eqvarp}). So for a shape-invariant intensity-dependent
nonresonant Jaynes-Cummings system, these matrix elements are given by

\begin{mathletters}
\label{eqs3p11id}
\begin{eqnarray}
\hat\sigma_{11}^P(t) &=& i{\gamma\over 2}\hat\nu_1^{-1}\sqrt{\hat
T\hat B_-}\left\{\hat z_2(t)\,{\cal
G}_{CS}^{(+)}(t;\hat\nu_2,\hat\omega_2,\hat\omega_1) -  \hat
y_2(t)\,{\cal G}_{SS}^{(+)}(t;\hat\nu_2,\hat\omega_2, \hat\omega_1)
\right\}\hat H_2^{3/4}\nonumber\\ &+& i{\gamma\over
2}\hat\nu_1^{-1}\hat H_2^{3/4}\left\{\hat z_1(t)\, {\cal
G}_{SC}^{(-)}(t;\hat\nu_1,\hat\omega_1,\hat\omega_2) -  \hat
y_1(t)\,{\cal G}_{CC}^{(-)}(t;\hat\nu_1,\hat\omega_1, \hat\omega_2)
\right\}\sqrt{\hat B_+\hat T^\dagger}\,,\\
\hat\sigma_{12}^P(t) &=& {\gamma\over 2}\hat\nu_1^{-1}\sqrt{\hat T\hat
B_-}\left\{\hat z_2(t)\,{\cal
G}_{CC}^{(+)}(t;\hat\nu_2,\hat\omega_2,\hat\omega_1) -  \hat
y_2(t)\,{\cal G}_{SC}^{(+)}(t;\hat\nu_2,\hat\omega_2, \hat\omega_1)
\right\}\sqrt{\hat H_2\hat T\hat B_-}\nonumber\\ &+& {\gamma\over
2}\hat\nu_1^{-1}\hat H_2^{3/4}\left\{\hat z_1(t)\, {\cal
G}_{SS}^{(-)}(t;\hat\nu_1,\hat\omega_1,\hat\omega_2) +  \hat
y_1(t)\,{\cal G}_{CS}^{(-)}(t;\hat\nu_1,\hat\omega_1, \hat\omega_2)
\right\}\hat H_1^{1/4}\,,\\
\hat\sigma_{21}^P(t) &=& {\gamma\over 2}\hat\nu_2^{-1}\sqrt{\hat
B_+\hat T^\dagger\hat H_2}\left\{\hat z_1(t)\,{\cal
G}_{CC}^{(+)}(t;\hat\nu_1,\hat\omega_1,\hat\omega_2) -  \hat
y_1(t)\,{\cal G}_{SC}^{(+)}(t;\hat\nu_1,\hat\omega_1, \hat\omega_2)
\right\}\sqrt{\hat B_+\hat T^\dagger}\nonumber\\ &+& {\gamma\over
2}\hat\nu_2^{-1}\hat H_1^{1/4}\left\{\hat z_2(t)\, {\cal
G}_{SS}^{(-)}(t;\hat\nu_2,\hat\omega_2,\hat\omega_1) -  \hat
y_2(t)\,{\cal G}_{CS}^{(-)}(t;\hat\nu_2,\hat\omega_2, \hat\omega_1)
\right\}\hat H_2^{3/4}\,,\\
\hat\sigma_{22}^P(t) &=& i{\gamma\over 2}\hat\nu_2^{-1}\sqrt{\hat
B_+\hat T^\dagger\hat H_2}\left\{\hat z_1(t)\,{\cal
G}_{CS}^{(+)}(t;\hat\nu_1,\hat\omega_1,\hat\omega_2) -  \hat
y_1(t)\,{\cal G}_{SS}^{(+)}(t;\hat\nu_1,\hat\omega_1, \hat\omega_2)
\right\}\hat H_1^{1/4}\nonumber\\ &+& i{\gamma\over
2}\hat\nu_2^{-1}\hat H_1^{1/4}\left\{\hat z_2(t)\, {\cal
G}_{SC}^{(-)}(t;\hat\nu_2,\hat\omega_2,\hat\omega_1) +  \hat
y_2(t)\,{\cal G}_{CC}^{(-)}(t;\hat\nu_2,\hat\omega_2, \hat\omega_1)
\right\}\sqrt{\hat H_2\hat T\hat B_-}\,.
\end{eqnarray}
\end{mathletters}
Yet the auxiliary functions, ${\cal G}^{(\pm)}_{XY}(t;\hat p,\hat
q,\hat r)$, are given by Eqs.~(\ref{eqfgpm}) and (\ref{eqfxy}). From
Eqs.~(\ref{eqsp0}) and (\ref{eqs3ini}), we have for the elements of the
population inversion matrix:

\begin{equation}
[\hat\sigma_3(t)]_{ij} = \cos{(\hat\nu_it)}\,
\left[\hat\sigma_3(0)\right]_{ij} +  {2i\alpha\over
\hbar}\,\sin{(\hat\nu_it)}\; \hat\nu_i^{-1} \left[\hat {\bf
S}_i(0)\,\hat\sigma_3(0)\right]_{ij} + \hat\sigma^P_{ij}(t)\,. 
\label{eqs3finid}
\end{equation}

\bigskip

{\bf a) The Intensity-Dependent Resonant Limit}
\bigskip

In this limit we set $(\Delta = 0)$, so the evolution matrix of the
system is given by

\begin{equation}
\hat {\bf U}_i(t,0) = \left[ \matrix{\cos{\left({1\over
2}\hat\nu_1t\right)} &  \sin{\left({1\over 2}\hat\nu_1 t\right)}\,\hat
C\cr -\sin{\left({1\over 2}\hat\nu_2 t\right)}\, \hat C^\dagger &
\cos{\left({1\over 2}\hat\nu_2 t\right)} \cr}\right]\,. 
\label{eqevo5id}
\end{equation}
and the elements of the population inversion factor can be written as

\begin{equation}
[\hat\sigma_3(t)]_{ij} = \cos{(\hat\nu_it)}\,
\left[\hat\sigma_3(0)\right]_{ij} +  {2i\alpha\over
\hbar}\,\sin{(\hat\nu_it)}\; \hat\nu_i^{-1} \left[\hat {\bf
S}_i(0)\,\hat\sigma_3(0)\right]_{ij}\,. 
\label{eqs3fresid}
\end{equation}

\bigskip

{\bf b) The Standard Intensity-Dependent Jaynes-Cummings Limit}
\bigskip

For the case of a harmonic oscillator system ($\hat T = \hat T^\dagger
\longrightarrow 1$, \ $\hat B_- \longrightarrow \hat a$, \ $\hat B_+
\longrightarrow \hat a^\dagger$ \ and \ $[\hat a,\hat a^\dagger] =
\hbar\omega$), we have for the $\hat\sigma_{ij}^P(t)$-matrix elements
the following expressions

\begin{mathletters}
\label{eqs3p11sid}
\begin{eqnarray}
\hat\sigma_{11}^P(t) &=& i{\gamma\over 2}\hat\nu_1^{-1} \sqrt{\hat
a}\left\{{\cal K}_S(t;\hat\omega_2,\hat\omega_1,\hat\nu_2) - {\cal
K}_S(t;\hat\omega_2,-\hat\omega_1,\hat\nu_2) \right\} \left(\hat a\hat
a^\dagger\right)^{3/4}\nonumber\\ &-& i{\gamma\over
2}\hat\nu_1^{-1}\left(\hat a\hat a^\dagger\right)^{3/4} \left\{{\cal
K}_S(t;\hat\omega_2,\hat\omega_1,\hat\nu_1) - {\cal
K}_S(t;\hat\omega_2,-\hat\omega_1,\hat\nu_1) \right\} \sqrt{\hat
a^\dagger }\\
\hat\sigma_{12}^P(t) &=& {\gamma\over 2}\hat\nu_1^{-1}\sqrt{\hat a}
\left\{{\cal K}_C(t;\hat\omega_2,\hat\omega_1,\hat\nu_2) - {\cal
K}_C(t;\hat\omega_2,-\hat\omega_1,\hat\nu_2)\right\} \sqrt{\hat a\hat
a^\dagger\hat a}\nonumber\\ &-& {\gamma\over
2}\hat\nu_1^{-1}\left(\hat a\hat a^\dagger\right)^{3/4} \left\{{\cal
K}_C(t;\hat\omega_2,\hat\omega_1,\hat\nu_1) - {\cal
K}_C(t;\hat\omega_2,-\hat\omega_1,\hat\nu_1)\right\} \left(\hat
a^\dagger\hat a\right)^{1/4}\\
\hat\sigma_{21}^P(t) &=& {\gamma\over 2}\hat\nu_2^{-1} \sqrt{\hat
a^\dagger\hat a\hat a^\dagger} \left\{{\cal
K}_C(t;\hat\omega_2,\hat\omega_1,\hat\nu_1) + {\cal
K}_C(t;\hat\omega_2,-\hat\omega_1,\hat\nu_1)\right\} \sqrt{\hat
a^\dagger}\nonumber\\ &-& {\gamma\over 2}\hat\nu_2^{-1}\left(\hat
a^\dagger\hat a\right)^{1/4} \left\{{\cal
K}_C(t;\hat\omega_2,\hat\omega_1,\hat\nu_2) - {\cal
K}_C(t;\hat\omega_2,-\hat\omega_1,\hat\nu_2)\right\} \left(\hat a\hat
a^\dagger\right)^{3/4}\\
\hat\sigma_{22}^P(t) &=& i{\gamma\over 2}\hat\nu_2^{-1} \sqrt{\hat
a^\dagger\hat a\hat a^\dagger }\left\{{\cal
K}_S(t;\hat\omega_2,\hat\omega_1,\hat\nu_1)  + {\cal
K}_S(t;\hat\omega_2,-\hat\omega_1,\hat\nu_1)\right\} \left(\hat
a^\dagger\hat a\right)^{1/4}\nonumber\\ &-& i{\gamma\over
2}\hat\nu_2^{-1}\left(\hat a^\dagger\hat a\right)^{1/4} \left\{{\cal
K}_S(t;\hat\omega_2,\hat\omega_1,\hat\nu_2) + {\cal
K}_S(t;\hat\omega_2,-\hat\omega_1,\hat\nu_2) \right\}\sqrt{\hat a\hat
a^\dagger\hat a}\,,
\end{eqnarray}
\end{mathletters}
where the auxiliary functions, ${\cal K}_S(t;\hat p,\hat q,\hat r)$
and ${\cal K}_C(t;\hat p,\hat q,\hat r)$, are given by
Eqs.~(\ref{eqksc}). The final expression for the population inversion
factor has the same form given by Eq.~(\ref{eqs3finid}) with

\begin{mathletters}
\label{eqfroscid}
\begin{eqnarray}
& & \hbar\hat \nu_1 = 2\alpha\;\hat a\hat a^\dagger
\,,\qquad\qquad\qquad \hbar\hat \nu_2 = 2\alpha\;\hat a^\dagger\hat
a\,,\\ & & \hbar\hat \omega_1 = \alpha\sqrt{(\hat a\hat a^\dagger)^2 +
\beta^2}\,,\qquad\qquad  \hbar\hat \omega_2 = \alpha\sqrt{(\hat
a^\dagger\hat a)^2 +\beta^2}\,. 
\end{eqnarray}
\end{mathletters}

\section{Conclusions}

In this article we extended our earlier work \cite{Aleixo:2000bd} on
bound-state problems which represent two-level systems. The
corresponding coupled-channel Hamiltonians generalize the nonresonant
and intensity-dependent nonresonant Jaynes-Cummings Hamiltonians. In
the case of a nonresonant system, if we take the starting Hamiltonian
to be the simplest shape-invariant system, namely the harmonic
oscillator, our results reduce to those of the standard nonresonant
Jaynes-Cummings approach, which has been extensively used to model a
two-level atom-single field mode interaction whose detuning it is not
null. In addition we have studied time evolution and population
inversion factor of the both kind of generalized systems.

These models are not only interesting on their own account. Being
exactly solvable coupled-channels problems they may help to assess the
validity and accuracy of various approximate approaches to the
coupled-channel problems \cite{Balantekin:1998yh}.

\section*{ACKNOWLEDGMENTS}

This work was supported in part by the U.S. National Science
Foundation Grants No.\ PHY-9605140 and PHY-0070161  at the University
of Wisconsin, and in part by the University of Wisconsin Research
Committee with funds granted by the Wisconsin Alumni Research
Foundation.   A.B.B.\ acknowledges the support of the Alexander von
Humboldt-Stiftung. M.A.C.R.\  acknowledges the support of Funda\c
c\~ao de Amparo \`a Pesquisa do Estado de S\~ao Paulo (Contract No.\
98/13722-2). A.N.F.A.  acknowledges the support of Funda\c c\~ao
Coordena\c c\~ao de Aperfei\c coamento de Pessoal de N\'{\i}vel
Superior (Contract No.  BEX0610/96-8). 

\bigskip\bigskip\bigskip\bigskip\bigskip

\appendix{\bf Appendix A}

\bigskip

Here we give the steps used to obtain the specific form of the
operators $\hat C$ and $\hat D$. Using Eq.~({\ref{eqevo3}) into the
unitary condition equation (\ref{equni}) actually we can show that the
$\hat C$ and $\hat D$ operators need to satisfy the following six
conditions 
\begin{mathletters}
\label{eqopcd}
\begin{eqnarray}
& & \hat C\hat C^\dagger = \hat C^\dagger\hat C = 1\\ & & \hat D\hat
D^\dagger = \hat D^\dagger\hat D = 1\\ & & \hat D^\dagger\,\sin{(\hat
\omega_2 t)} = - \sin{(\hat \omega_1 t)}\;\hat C\\ & & \hat
D\,\cos{(\hat \omega_1 t)} =  -\cos{(\hat \omega_2 t)}\;\hat
C^\dagger\,. 
\end{eqnarray}
\end{mathletters}
At this point we can use the following property 
\begin{eqnarray}
\sqrt{\hat T\hat B_-}\;\hat\omega_2 &=&  \sqrt{\hat T\hat
B_-}\;\sqrt{\alpha^2\hat B_+\hat B_-+\beta^2}/\hbar\nonumber\\ &=&
\sqrt{\alpha^2\hat T\hat B_-\hat B_+\hat B_-+\hat T\hat
B_-\beta^2}/\hbar\nonumber\\ &=& \sqrt{\alpha^2\hat T\hat B_-\hat
B_+\hat T^\dagger\hat T\hat B_-+\beta^2\hat T\hat
B_-}/\hbar\nonumber\\ &=& \sqrt{\alpha^2\hat T\hat B_-\hat B_+\hat
T^\dagger +\beta^2}/\hbar\;\sqrt{\hat T\hat B_-}\nonumber\\ &=&
\hat\omega_1\;\sqrt{\hat T\hat B_-}\,. 
\label{eqomth1}
\end{eqnarray}
Then, with this result we have 
\begin{equation}
\sqrt{\hat T\hat B_-}\;\hat\omega_2^2  = \sqrt{\hat T\hat
B_-}\;\hat\omega_2\;\hat\omega_2 = \hat\omega_1\;\sqrt{\hat T\hat
B_-}\;\hat\omega_2 = \hat\omega_1^2\;\sqrt{\hat T\hat B_-}\,,
\label{eqomth2}
\end{equation}
and finally, by induction, we conclude that 
\begin{equation}
\sqrt{\hat T\hat B_-}\;\hat\omega_2^n =  \hat\omega_1^n\;\sqrt{\hat
T\hat B_-}\,. 
\label{eqomth}
\end{equation}
In the same way,  
\begin{eqnarray}
\sqrt{\hat B_+\hat T^\dagger}\;\hat\omega_1 &=&  \sqrt{\hat B_+\hat
T^\dagger}\;\sqrt{\alpha^2\hat T\hat B_-\hat B_+\hat T^\dagger +
\beta^2}/\hbar\nonumber\\ &=& \sqrt{\alpha^2\hat B_+\hat T^\dagger\hat
T\hat B_-\hat B_+\hat T^\dagger + \hat B_+\hat
T^\dagger\beta^2}/\hbar\nonumber\\ &=& \sqrt{\alpha^2\hat B_+\hat
B_-\hat B_+\hat T^\dagger + \beta^2\hat B_+\hat T^\dagger
}/\hbar\nonumber\\ &=& \sqrt{\alpha^2\hat B_+\hat B_-
+\beta^2}/\hbar\;\sqrt{\hat B_+\hat T^\dagger }\nonumber\\ &=&
\hat\omega_2\;\sqrt{\hat B_+\hat T^\dagger}\,. 
\label{eqomth12}
\end{eqnarray}
Then, with this result we have 
\begin{equation}
\sqrt{\hat B_+\hat T^\dagger }\;\hat\omega_1^2    = \sqrt{\hat B_+\hat
T^\dagger }\;\hat\omega_1\;\hat\omega_1  = \hat\omega_2\;\sqrt{\hat
B_+\hat T^\dagger}\;\hat\omega_1 = \hat\omega_2^2\;\sqrt{\hat B_+\hat
T^\dagger}\,,
\label{eqomth22}
\end{equation}
and finally, again by induction, we get 
\begin{equation}
\sqrt{\hat B_+\hat T^\dagger }\;\hat\omega_1^n =
\hat\omega_2^n\;\sqrt{\hat B_+\hat T^\dagger}\,. 
\label{eqomthx}
\end{equation} 
Using the properties given by Eqs.~(\ref{eqomth}) and (\ref{eqomthx})
and the forms of $\hat C$, $\hat D$ operators, defined by
Eqs.~(\ref{eqcdfin}), we can verify that 
\begin{equation}
\hat C\hat C^\dagger = \hat D^\dagger\hat D  = {i\over \hat
H_2^{1/4}}\sqrt{\hat T\hat B_-}\;\sqrt{\hat B_+\hat
T^\dagger}{(-i)\over \hat H_2^{1/4}} = {1\over \hat
H_2^{1/4}}\sqrt{\hat H_2}{1\over \hat H_2^{1/4}} = 1\,,
\label{eqcd1}
\end{equation}
and 
\begin{equation}
\hat C^\dagger\hat C = \hat D\hat D^\dagger  = \sqrt{\hat B_+\hat
T^\dagger}{(-i)\over \hat H_2^{1/4}}\;{i\over \hat
H_2^{1/4}}\sqrt{\hat T\hat B_-} = \sqrt{\hat B_+\hat T^\dagger}{1\over
\sqrt{\hat H_2}}\;\sqrt{\hat H_2}{1\over \sqrt{\hat B_+\hat
T^\dagger}} = 1\,.
\label{eqcd2}
\end{equation}
Also using the series expansion of the trigonometric functions, we can
show that
\begin{eqnarray}
\hat D^\dagger \,\sin{(\hat\omega_2t)} &=& {-i\over\hat
H_2^{1/4}}\sqrt{\hat T\hat B_-}\sum_{n=0}^\infty (-1)^n
{\left(\hat\omega_2t\right)^{2n+1}\over (2n+1)!}\nonumber\\ &=&
{-i\over\hat H_2^{1/4}}\sum_{n=0}^\infty (-1)^n \sqrt{\hat T\hat
B_-}{\left(\hat\omega_2t\right)^{2n+1}\over (2n+1)!}\nonumber\\ &=&
{-i\over\hat H_2^{1/4}}\sum_{n=0}^\infty (-1)^n
{\left(\hat\omega_1t\right)^{2n+1} \over (2n+1)!}\sqrt{\hat T\hat
B_-}\nonumber\\ &=& \sum_{n=0}^\infty (-1)^n
{\left(\hat\omega_1t\right)^{2n+1} \over (2n+1)!}{-i\over\hat
H_2^{1/4}}\sqrt{\hat T\hat B_-}\nonumber\\ &=&
-\sin{(\hat\omega_1t)}\,\hat C\,,
\label{eqcsin}
\end{eqnarray}
where we used the commutation between \ $\hat H_2$ \ and \
$\hat\omega_1$ (see Appendix B). In the same way we can prove that 
\begin{eqnarray}
\hat D \,\cos{(\hat\omega_1t)} &=& \sqrt{\hat B_+\hat T^\dagger}
{i\over\hat H_2^{1/4}}\sum_{n=0}^\infty (-1)^n
{\left(\hat\omega_1t\right)^{2n}\over (2n)!}\nonumber\\ &=&
\sum_{n=0}^\infty (-1)^n\sqrt{\hat B_+\hat T^\dagger}
{\left(\hat\omega_1t\right)^{2n}\over (2n)!}{i\over\hat H_2^{1/4}}
\nonumber\\ &=& \sum_{n=0}^\infty (-1)^n
{\left(\hat\omega_2t\right)^{2n}\over (2n)!}\sqrt{\hat B_+\hat
T^\dagger} {i\over\hat H_2^{1/4}}\nonumber\\ &=&
-\cos{(\hat\omega_2t)}\,\hat C^\dagger\,. 
\label{eqdcos}
\end{eqnarray}
Again, we used the commutation between \ $\hat H_2$ \ and \
$\hat\omega_1$.


\bigskip\bigskip\bigskip

\appendix{\bf Appendix B}

\bigskip

In this appendix we show the necessary steps to obtain the explicit
expressions of the particular solution elements of the population
inversion factor. To resolve the integrals in Eq.~(\ref{eqvarp}),
first we need to determine the elements of the \ $\hat {\bf
F}(t)$-matrix.  To do that we can use Eqs.~(\ref{eqso}),
(\ref{eqomft}),  and (\ref{eqevo4}) to write down 
\begin{mathletters}
\label{eqlfelem}
\begin{eqnarray}
\hat F_{11}(t) &=& -\gamma\left\{\cos{(\hat\omega_1t)}\,\hat T\hat
B_-\,\sin{(\hat\omega_2t)}\,\hat C^\dagger + \hat
C\,\sin{(\hat\omega_2t)}\,\hat B_+\hat
T^\dagger\,\cos{(\hat\omega_1t)}\right\}\nonumber\\ &=&
i\gamma\left\{\sqrt{\hat T\hat
B_-}\,\cos{(\hat\omega_2t)}\,\sin{(\hat\omega_1t)}\,\hat H_2^{1/4} -
\hat
H_2^{1/4}\,\sin{(\hat\omega_1t)}\,\cos{(\hat\omega_2t)}\,\sqrt{\hat
B_+\hat T^\dagger}\right\}\\ \hat F_{12}(t) &=&
\gamma\left\{\cos{(\hat\omega_1t)}\,\hat T\hat
B_-\,\cos{(\hat\omega_2t)} - \hat C\,\sin{(\hat\omega_2t)}\,\hat
B_+\hat T^\dagger\,\sin{(\hat\omega_1t)}\,\hat C\right\}\nonumber\\
&=& \gamma\left\{\sqrt{\hat T\hat
B_-}\,\cos{(\hat\omega_2t)}\,\cos{(\hat\omega_1t)}\,\sqrt{\hat T\hat
B_-} + \hat H_2^{1/4}\,\sin{(\hat\omega_1t)}\,\sin{(\hat\omega_2t)}\,
\hat H_1^{1/4}\right\}\\ \hat F_{21}(t) &=&
\gamma\left\{\cos{(\hat\omega_2t)}\,\hat B_+\hat
T^\dagger\,\cos{(\hat\omega_1t)} - \hat
C^\dagger\,\sin{(\hat\omega_1t)}\,\hat T\hat
B_-\,\sin{(\hat\omega_2t)}\,\hat C^\dagger\right\}\nonumber\\ &=&
\gamma\left\{\sqrt{\hat B_+\hat
T^\dagger}\,\cos{(\hat\omega_1t)}\,\cos{(\hat\omega_2t)}\,\sqrt{\hat
B_+\hat T^\dagger} + \hat
H_1^{1/4}\,\sin{(\hat\omega_2t)}\,\sin{(\hat\omega_1t)}\, \hat
H_2^{1/4}\right\}\\ \hat F_{22}(t) &=& \gamma\left\{\hat
C^\dagger\,\sin{(\hat\omega_1t)}\, \hat T\hat
B_-\,\cos{(\hat\omega_2t)} + \cos{(\hat\omega_2t)}\,\hat B_+\hat
T^\dagger\,\sin{(\hat\omega_1t)}\,\hat C\right\}\nonumber\\ &=&
i\gamma\left\{\sqrt{\hat B_+\hat
T^\dagger}\,\cos{(\hat\omega_1t)}\,\sin{(\hat\omega_2t)}\,\hat
H_1^{1/4} -  \hat
H_1^{1/4}\,\sin{(\hat\omega_2t)}\,\cos{(\hat\omega_1t)}\,\sqrt{\hat
T\hat B_-}\right\}\,,
\end{eqnarray}
\end{mathletters}
where \ $\gamma = 4\alpha^2\beta/\hbar^2$.
Here we used the properties (\ref{eqopcd}), (\ref{eqomth}) and
(\ref{eqomthx}), together with the following operators relations 
\begin{mathletters}
\label{eqprch}
\begin{eqnarray}
& & \hat C\,\sqrt{\hat B_+\hat T^\dagger} = - \sqrt{\hat T\hat
B_-}\,\hat C^\dagger = i\hat H_2^{1/4}\\ & & \sqrt{\hat B_+\hat
T^\dagger}\,\hat C  = - \hat C^\dagger  \,\sqrt{\hat T\hat B_-}  =
i\hat H_1^{1/4}\,. 
\end{eqnarray}
\end{mathletters} 
Now, keeping in mind that \  $\left[\hat\nu_j,\hat\omega_j\right] =
0$,  $(j = 1,\, {\rm or}\; 2)$, so we may use the trigonometric
relationships  involving products of trigonometric functions with
arguments $\hat\nu_jt$ and $\hat\omega_jt$ (since we have \
$\exp{(\hat\nu_jt)}\,\exp{(\pm\hat\omega_jt)} =
\exp{[(\hat\nu_j\pm\hat\omega_j)t]}$).  Then, using those
relationships, the following commutators 
\begin{equation}
\left[\hat\nu_1,\hat H_2\right] = \left[\hat\omega_1,\hat H_2\right] =
\left[\hat\nu_2,\hat H_1\right] = \left[\hat\omega_2,\hat H_1\right] =
0\,,
\label{eqcomm}
\end{equation}
and the same properties (\ref{eqopcd}), (\ref{eqomth}) and
(\ref{eqomthx}), we can show that 
\begin{mathletters}
\label{eqlyfele}
\begin{eqnarray}
\hat y_1(t)\,\hat F_{11}(t) &=&  i{\gamma\over 2}\sqrt{\hat T\hat
B_-}\left\{\cos{\left[(\hat\nu_2-\hat\omega_2)t\right]}\,
\sin{(\hat\omega_1t)} + \cos{\left[(\hat\nu_2+\hat\omega_2)t\right]}\,
\sin{(\hat\omega_1t)}\right\}\hat H_2^{1/4}\nonumber\\ &+&
i{\gamma\over 2}\hat
H_2^{1/4}\left\{\sin{\left[(\hat\nu_1-\hat\omega_1)t\right]}\,
\cos{(\hat\omega_2t)} - \sin{\left[(\hat\nu_1+\hat\omega_1)t\right]}\,
\cos{(\hat\omega_2t)} \right\}\sqrt{\hat B_+\hat T^\dagger}\\
\hat y_1(t)\,\hat F_{12}(t) &=&  {\gamma\over 2}\sqrt{\hat T\hat
B_-}\left\{\cos{\left[(\hat\nu_2-\hat\omega_2)t\right]}\,
\cos{(\hat\omega_1t)} + \cos{\left[(\hat\nu_2+\hat\omega_2)t\right]}\,
\cos{(\hat\omega_1t)}\right\}\sqrt{\hat T\hat B_-} \nonumber\\ &+&
{\gamma\over 2}\hat
H_2^{1/4}\left\{\sin{\left[(\hat\nu_1+\hat\omega_1)t\right]}\,
\sin{(\hat\omega_2t)} - \sin{\left[(\hat\nu_1-\hat\omega_1)t\right]}\,
\sin{(\hat\omega_2t)} \right\}\hat H_1^{1/4}\\
\hat y_2(t)\,\hat F_{21}(t) &=&  {\gamma\over 2}\sqrt{\hat B_+\hat
T^\dagger}\left\{\cos{\left[(\hat\nu_1-\hat\omega_1)t\right]}\,
\cos{(\hat\omega_2t)} + \cos{\left[(\hat\nu_1+\hat\omega_1)t\right]}\,
\cos{(\hat\omega_2t)}\right\}\sqrt{\hat B_+\hat T^\dagger} \nonumber\\
&+& {\gamma\over 2}\hat
H_1^{1/4}\left\{\sin{\left[(\hat\nu_2+\hat\omega_2)t\right]}\,
\sin{(\hat\omega_1t)} - \sin{\left[(\hat\nu_2-\hat\omega_2)t\right]}\,
\sin{(\hat\omega_1t)} \right\}\hat H_2^{1/4}\\
\hat y_2(t)\,\hat F_{22}(t) &=&  i{\gamma\over 2}\sqrt{\hat B_+\hat
T^\dagger}\left\{\cos{\left[(\hat\nu_1-\hat\omega_1)t\right]}\,
\sin{(\hat\omega_2t)} + \cos{\left[(\hat\nu_1+\hat\omega_1)t\right]}\,
\sin{(\hat\omega_2t)}\right\}\hat H_1^{1/4}\nonumber\\ &+&
i{\gamma\over 2}\hat
H_1^{1/4}\left\{\sin{\left[(\hat\nu_2-\hat\omega_2)t\right]}\,
\cos{(\hat\omega_1t)} - \sin{\left[(\hat\nu_2+\hat\omega_2)t\right]}\,
\cos{(\hat\omega_1t)} \right\}\sqrt{\hat T\hat B_-} \,. 
\end{eqnarray}
\end{mathletters}
In a similar way, we can show that 
\begin{mathletters}
\label{eqlzfele}
\begin{eqnarray}
\hat z_1(t)\,\hat F_{11}(t) &=&  i{\gamma\over 2}\sqrt{\hat T\hat
B_-}\left\{\sin{\left[(\hat\nu_2-\hat\omega_2)t\right]}\,
\sin{(\hat\omega_1t)} + \sin{\left[(\hat\nu_2+\hat\omega_2)t\right]}\,
\sin{(\hat\omega_1t)}\right\}\hat H_2^{1/4}\nonumber\\ &-&
i{\gamma\over 2}\hat
H_2^{1/4}\left\{\cos{\left[(\hat\nu_1-\hat\omega_1)t\right]}\,
\cos{(\hat\omega_2t)} - \cos{\left[(\hat\nu_1+\hat\omega_1)t\right]}\,
\cos{(\hat\omega_2t)} \right\}\sqrt{\hat B_+\hat T^\dagger}\\
\hat z_1(t)\,\hat F_{12}(t) &=&  {\gamma\over 2}\sqrt{\hat T\hat
B_-}\left\{\sin{\left[(\hat\nu_2-\hat\omega_2)t\right]}\,
\cos{(\hat\omega_1t)} + \sin{\left[(\hat\nu_2+\hat\omega_2)t\right]}\,
\cos{(\hat\omega_1t)}\right\}\sqrt{\hat T\hat B_-} \nonumber\\ &-&
{\gamma\over 2}\hat
H_2^{1/4}\left\{\cos{\left[(\hat\nu_1+\hat\omega_1)t\right]}\,
\sin{(\hat\omega_2t)} - \cos{\left[(\hat\nu_1-\hat\omega_1)t\right]}\,
\sin{(\hat\omega_2t)} \right\}\hat H_1^{1/4}\\
\hat z_2(t)\,\hat F_{21}(t) &=&  {\gamma\over 2}\sqrt{\hat B_+\hat
T^\dagger}\left\{\sin{\left[(\hat\nu_1-\hat\omega_1)t\right]}\,
\cos{(\hat\omega_2t)} + \sin{\left[(\hat\nu_1+\hat\omega_1)t\right]}\,
\cos{(\hat\omega_2t)}\right\}\sqrt{\hat B_+\hat T^\dagger} \nonumber\\
&-& {\gamma\over 2}\hat
H_1^{1/4}\left\{\cos{\left[(\hat\nu_2+\hat\omega_2)t\right]}\,
\sin{(\hat\omega_1t)} - \cos{\left[(\hat\nu_2-\hat\omega_2)t\right]}\,
\sin{(\hat\omega_1t)} \right\}\hat H_2^{1/4}\\
\hat z_2(t)\,\hat F_{22}(t) &=&  i{\gamma\over 2}\sqrt{\hat B_+\hat
T^\dagger}\left\{\sin{\left[(\hat\nu_1-\hat\omega_1)t\right]}\,
\sin{(\hat\omega_2t)} + \sin{\left[(\hat\nu_1+\hat\omega_1)t\right]}\,
\sin{(\hat\omega_2t)}\right\}\hat H_1^{1/4}\nonumber\\ &-&
i{\gamma\over 2}\hat
H_1^{1/4}\left\{\cos{\left[(\hat\nu_2-\hat\omega_2)t\right]}\,
\cos{(\hat\omega_1t)} - \cos{\left[(\hat\nu_2+\hat\omega_2)t\right]}\,
\cos{(\hat\omega_1t)} \right\}\sqrt{\hat T\hat B_-} \,. 
\end{eqnarray}
\end{mathletters}

The non-commutativity between the operators $\hat\omega_1$ and
$\hat\omega_2$ imply that to calculate the integrals involving the
terms given by Eqs.~(\ref{eqlyfele}) and (\ref{eqlzfele}) we need  to
use the series expansion of the trigonometric functions. In this  case
the integrals can be easily done because the time variable can be
considered as a parameter factor. Finally, using these results into
Eq.~(\ref{eqvarp}) is trivial to find the expression  (\ref{eqs3p11})
for the matrix elements of the particular solution. 


\end{document}